\documentclass[iop]{emulateapj}

\usepackage{bm}
\usepackage{float}

\newcommand{\msunyr}{M_{\odot}\:{\rm yr}^{-1}}
\newcommand{\mdot}{\dot{m}}
\newcommand{\msun}{M_{\odot}}
\newcommand{\rsun}{R_{\odot}}
\newcommand{\lsun}{L_{\odot}}

\newcommand{\cloudy}{C{\small LOUDY}}
\newcommand{\HII}{\ion{H}{2}}
\newcommand{\Sigcl}{\Sigma_{\rm cl}}
\newcommand{\gcm}{{\rm g\:cm^{-2}}}
\newcommand{\kms}{{\rm\:km\:s^{-1}}}

\shorttitle{Outflow-Confined HII Regions}
\shortauthors{Tanaka, Tan, \& Zhang}

\begin{document}

\title{Outflow-Confined \ion{H}{2} Regions. I. First Signposts of Massive Star Formation}

\author{Kei E. I. Tanaka}
\affil{Department of Astronomy, University of Florida, Gainesville, FL 32611, USA; ktanaka@ufl.edu}
\author{Jonathan C. Tan}
\affil{Departments of Astronomy \& Physics, University of Florida, Gainesville, FL 32611, USA; jt@astro.ufl.edu}
\and
\author{Yichen Zhang}
\affil{Departamento de Astronom\'ia, Universidad de Chile, Santiago, Chile; yzhang@das.uchile.cl}

\begin{abstract}
We present an evolutionary sequence of models of the photoionized
disk-wind outflow around forming massive stars based on the Core
Accretion model.  The outflow is expected to be the first structure to
be ionized by the protostar and can confine the expansion of the
\HII~region, especially in lateral directions in the plane of the
accretion disk. The ionizing luminosity increases as Kelvin-Helmholz
contraction proceeds, and the \HII~region is formed when the stellar
mass reaches $\sim10\:$ -- $\:20\msun$ depending on the initial cloud
core properties.  Although some part of outer disk surface remains
neutral due to shielding by the inner disk and the disk wind, almost
the whole of the outflow is ionized in $10^3$--$10^4{\rm yr}$ after
initial \HII~region formation. Having calculated the extent and
temperature structure of the \HII~region within the immediate
protostellar environment, we then make predictions for the strength of
its free-free continuum and recombination line emission.
The free-free radio emission from the ionized outflow has a flux density of
$\sim (20\:$--$\:200) \times (\nu/10 {\rm GHz})^p\:{\rm mJy}$
for a source at a distance of 1 kpc
with a spectral index $p\simeq0.4\:$--$\:0.7$, and the apparent size is
typically $\sim$500~AU at 10~GHz.  The $\rm H40\alpha$
line profile has a width of about $100\:\kms$.  These properties of
our model are consistent with observed radio winds and jets around
forming massive protostars.
\end{abstract}

\keywords{stars: formation, evolution - accretion, accretion disks.}

\section{Introduction}

The formation mechanism of massive stars remains much debated with
Core Accretion, Competitive Accretion and even Protostellar Collisions
still discussed \citep[see][for a recent review]{tan14}.

One of the key differences between high and low-mass protostars is
that the former are expected to complete the later stages of their
accretion while fusing hydrogen on the main sequence. Their
photospheres should be hot and strong sources of Lyman continnum
photons with energies $>13.6$~eV, which can photoionize H to create an
\ion{H}{2} region. These protostellar \ion{H}{2} regions are important
for at least two main reasons. First, by their long wavelength radio
free-free and Hydrogen recombination line (HRL) emission they provide
diagnostics of the massive star formation process. Second, the ionized
gas reaches temperatures of $\sim 10^4$~K and its thermal pressure
could be an important feedback process that regulates massive star
formation, either by disrupting the wider scale accretion reservoir
\citep[i.e., core envelope or Bondi-Hoyle competitive accretion
  region: e.g.,][]{dal05,pet10,pet11} or by photoevaporation of the
accretion disk \citep{hol94,yor96,ric97,mck08,tan13}.

Our goal in this study is to present model calculations of the
structure of the earliest-stage protostellar \ion{H}{2} regions in the
context of the Turbulent Core Model of \citet{mck03} (hereafter,
MT03).  Early work in this direction was carried out by \citet{tan03},
who described these structures as ``Outflow-Confined'' \ion{H}{2}
regions, since the magneto-centrifugally-launched protostellar outflow
(either disk-wind, X-wind or both) will be the first barrier to the
ionizing photons. 

\citet{ket07} studied the structure and evolution of \HII~regions that
are confined by the infalling flow onto the protostellar disk.
However, their model did not include the effect of the MHD-driven
outflow. If massive stars form in a scaled-up, but similar manner to
low-mass stars, then we expect such outflows to always be present
around massive protostars during their main accretion phase. Indeed,
larger-scale collimated outflows are often observed to be driven from
massive protostars
\citep[e.g.,][]{beu02,gar07,lop09}

A series of massive protostellar evolution calculations to predict
continuum radiative transfer, especially infrared to sub-mm
morphologies and spectral energy distributions (SEDs), has been
carried out by \citet{zha11}, \citet{zha13}, and \citet{zha14}
(hereafter, ZTH14).  Here we will calculate the ionization structure
and associated radio emission predicted by these protostellar
evolutionary sequences.

Study of radio emission from massive protostars has a long history
\citep[see, e.g.,][]{woo89,hoa07,tan14}.  Ultra-compact
($\leq0.1\:{\rm pc}$) and, especially, hyper-compact ($\leq0.01\:{\rm
  pc}$) \ion{H}{2} regions may include sources that are in the
protostellar phase. Elongated radio continuum sources, e.g., radio
``jets'', have been seen in a number of sources
\citep[e.g.,][]{guz12}. The eventual goal of our work will be to
compare the developed models with such observations, however, detailed
comparison with individual sources, requiring tailored protostellar
models, is deferred to future works in this series.

This paper is organized as follows.  In \S\ref{sec_method} we describe
our methods: in particular, we briefly review the protostellar
evolution models, describe calculation of the photoionized structure
and the radiative transfer of free-free continuum and HRL emission.
Next, in \S\ref{sec_results}, we present our results: especially, we
explore the evolution of the photoionized regions for various initial
core conditions and predict their observational features. In
\S\ref{sec_discussion} we discuss the implications of our results,
especially summarizing the evolution of the outflow confined
\HII~regions, and making an initial comparison of our models with the
general characteristics of observed HC/UC \HII~regions and radio
winds/jets.  We conclude in \S\ref{sec_conclusions}.

\section{Methods}\label{sec_method}

\subsection{Evolution of protostars, disks, envelopes and outflows} \label{sec_model}

\begin{deluxetable*}{ccc|ccccc}
\tablecolumns{8}
\tablewidth{0pc}
\tablecaption{Model lists  \label{tab_model}}
\startdata
\hline \hline
Models & $M_{\rm c}\:(\msun)$ & $\Sigma_{\rm cl}\:({\rm g~cm^{-2}})$
& $m_*\:(\msun)$ & $r_*\:(\rsun)$
& $T_{\rm *,acc}\:({\rm \times 10^4 K})$ & $L_{\rm *,acc}\:(\times 10^4\lsun)$
& $\log_{10} S_{*,\rm acc}\:({\rm~s^{-1}})$ \\
\hline \hline
A12 & 60 & 1 & 12 & 12.9 & 2.33 & 4.4 & 46.5 \\
A16 &      &    & 16 & 6.4   & 3.67 & 6.6 & 48.3 \\
A24 &      &    & 24 & 6.4   & 3.90 & 8.5 & 48.6 \\ \hline
Al08 & 60 & 0.316 &   8 & 14.0 & 1.59 & 1.1 & 43.9 \\
Al12 &      &           & 12 & 4.7   & 3.27 & 2.3 & 47.5\\
Al16 &      &           & 16 & 5.1   & 3.43 & 3.2 & 47.8 \\ \hline
Ah12 & 60 &  3.16  & 12 & 50.2 & 1.21 & 4.9 & 42.5 \\
Ah16 &      &           & 16 & 26.1 & 2.05 & 10.9 & 46.1 \\
Ah24 &      &           & 24 & 9.0   & 4.05 & 19.6 & 48.9 \\ \hline
B12 & 120 & 1  & 12 & 17.1 & 2.04 & 4.6 & 45.7 \\
B16 &        &     & 16 & 8.4 & 3.43 & 8.7 & 48.3 \\
B24 &        &     & 24 & 6.6 & 4.20 & 12.1 & 48.8 \\
B32 &        &     & 32 & 7.6 & 4.36 & 18.9 & 49.0 \\ \hline
C12 & 240 & 1  & 12 & 21.6 & 1.80 & 4.4 & 45.2 \\
C16 &        &     & 16 & 11.1 & 3.04 & 9.6 & 48.1 \\
C24 &        &     & 24 & 6.8 & 4.30 & 14.3 & 48.9 \\
C32 &        &     & 32 & 7.7 & 4.50 & 22.1 & 49.1 \\ 
C48 &        &     & 48 & 9.6 & 4.76 & 43.1 & 49.4 \\
C64 &        &     & 64 & 11.5 & 4.91 & 68.4 & 49.6
\enddata
\end{deluxetable*}

\begin{figure}
\begin{center}
\includegraphics[width=90mm]{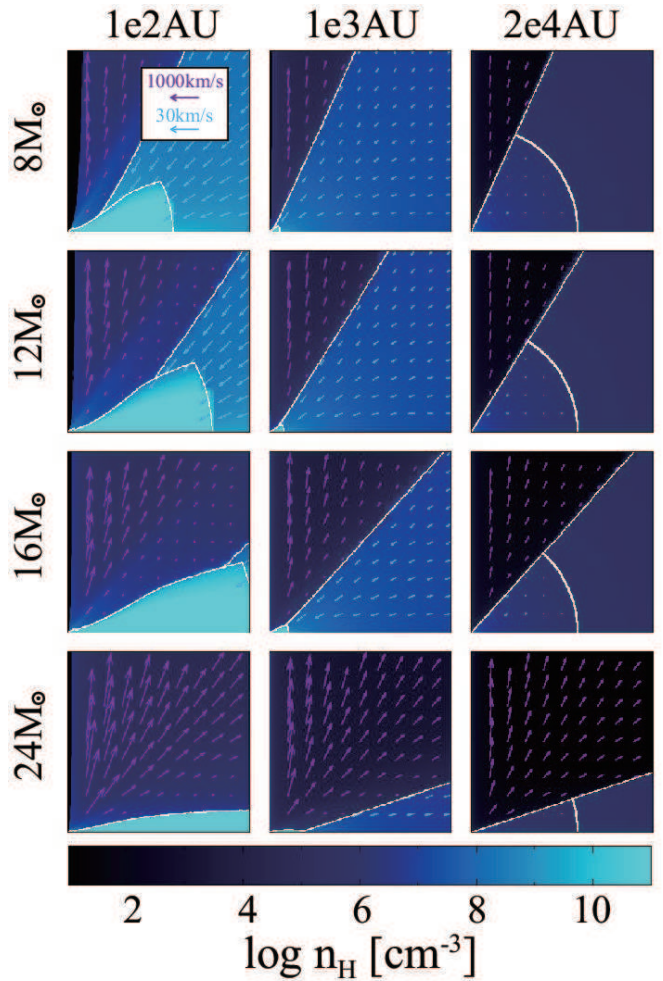}
\end{center}
\caption{
Evolutionary sequence of density, $n_{\rm H}$, and velocity fields
around protostars of mass $m_*=8,~12,~16,{\rm~and~}24M_\odot$ forming
from a core with initial properties $M_{c}=60M_\odot,~\Sigma_{\rm
  cl}=1\:{\rm g~cm^{-2}},$ and $\beta_{c}=0.02$ (model A08, A12, A16,
and A24).  Three spatial scales, successively zooming out from the
protostar, are shown from left to right.  In each panel, the protostar
is at the lower left corner, the horizontal axis lies on the disk
midplane and the vertical axis along the outflow axis.  The velocity
fields are shown by arrows (purple for the outflow; light-blue for the
infalling envelope; note the different velocity scales).
The initial core radius, i.e., its boundary with the clump,
and the boundaries of disk, envelope and outflow are indicated by white lines.}
\label{fig_n}
\end{figure}

\begin{figure}
\begin{center}
\includegraphics[width=90mm]{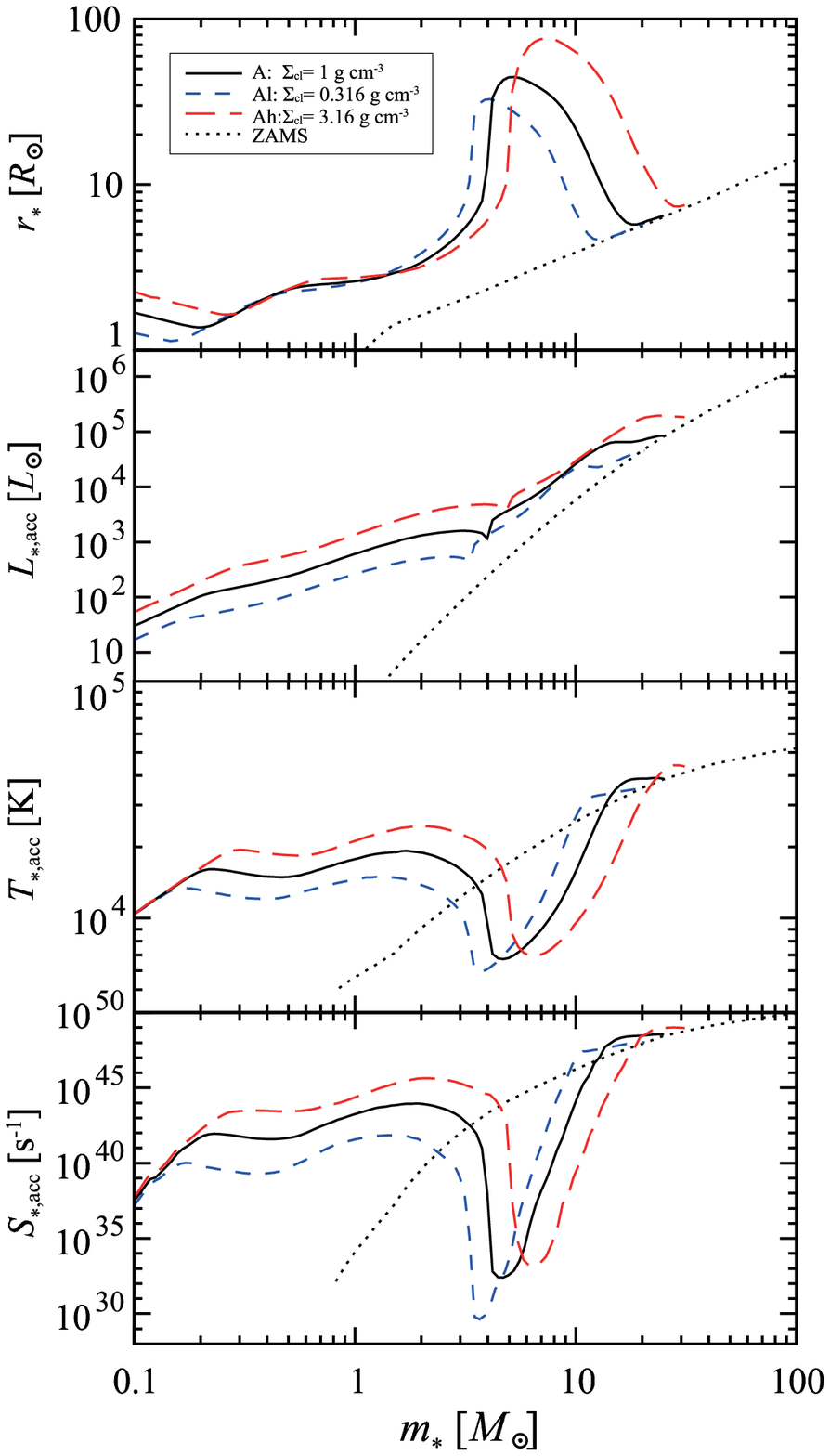}
\end{center}
\caption{
Evolutionary sequences of $r_{*}$, $L_{*,\rm acc}$, $T_{*,\rm acc}$,
and $S_{*,\rm acc}$ versus protostellar mass (top to bottom) for
protostars forming from $M_{c}=60M_\odot$ and $~\beta_{\rm
  c}=0.02$. The black solid line shows the case with $\Sigma_{\rm
  cl}=1\:{\rm g~cm^{-2}}$, while blue dashed and red long-dashed lines
show the $\Sigma_{\rm cl}=0.316$ and $3.16\:{\rm g~cm^{-2}}$ cases,
respectively.  The ZAMS properties are shown by the dotted lines.}
\label{fig_protost}
\end{figure}

We calculate the photoionization structure and present synthetic
observations in the context of the Turbulent Core Accretion Model of
MT03. We use a series of models of protostellar evolution and
surrounding gas structure constructed by ZTH14, which
self-consistently include the evolution of the protostar and disk
formed from a rotating parent core embedded in a particular clump
  environment, and mass accretion regulated by disk-wind outflows. We
note that these models are highly idealized in that they assume
axisymmetry about a single protostar, while real objects will be more
irregular, especially if the initial core is highly turbulent, and may
be forming in a cluster environment with other protostars in the
vicinity. Even if there is a relatively ordered and isolated core that
is collapsing to a central disk, this collapse may lead to binary or
low-order multiple star formation. Still, by first presenting the
simplest case of formation of an isolated massive protostar, we can
then see how relevant such models are as approximations to real
systems.

Here we briefly review the evolution of the protostar and its
surrounding gas structures. The initial core is assumed to be
spherical and in near virial equilibrium via support by turbulence
and/or magnetic fields.  The evolutionary track of a core is
determined by three environmental initial conditions: core mass
$M_{\rm c}$, mass surface density of ambient clump $\Sigcl$ (which
sets surface pressure of the core), and ratio of the core's initial
rotational to gravitational energy $\beta_{c}$.  We fix the rotational
parameter $\beta_{c} = 0.02$ as a typical value from observations of
lower-mass cores \citep{goo93,li12,pal13}.  The core density structure
is described by a power law in spherical radius, $\rho\propto
r^{-k_{\rho}}$. Values of $k_{\rho}\simeq1.3$--$1.6$ are inferred from
observations of dense cores in Infrared Dark Clouds (IRDCs)
\citep{but12,but14}.  We adopt $k_{\rho}=1.5$ as a fiducial value,
following MT03. Then, the radius of a core is $R_{c}=1.2\times10^4
(M_{c}/60\msun)^{1/2}(\Sigcl/1~\gcm)^{-1/2}{\rm AU}$.  Since the core
is in pressure equilibrium with the surrounding clump, it is more
compact for higher $\Sigcl$. 

Table \ref{tab_model} presents the parameter sets of $(M_c,~\Sigcl)$
for which we study the ionization structures. These are the same as
studied by ZTH14.  We refer to model group A as a fiducial group with
$M_{c}=60\msun$ and $\Sigcl=1~\gcm$.  We also investigate high- and
low-$\Sigcl$ cases, model groups Al and Ah, and $M_{c}=120\:\msun$ and
240$\msun$ cases, as model groups B and C, respectively.

The core undergoes inside-out collapse \citep{shu77,mcl97} with the
effect of rotation described with the solution by \citet{ulr76}
\citep[see also][]{tan04}. Given the high accretion rates associated
with collapse of dense cores, the protostellar disk is expected to be
massive and self-gravitating such that angular momentum may be
transported by torques due to, e.g., spiral arm structures.
Considering this efficient angular momentum transfer, the mass ratio
between disk and protostar is set to be a constant
$f_{d}=m_{d}/m_*=1/3$ \citep[e.g.,][]{kra08}. The disk structure is
described by a standard $\alpha$-viscosity disk model \citep{sha73},
but also including the effects of infall and outflow (conservations of
mass and angular momentum). The temperatures at the midplane are
evaluated by vertical energy balance, and the vertical structure of
the disk is determined by this temperature. The disk surface is set
by assuming continuity with the density structure of the outflow.
More details about the disk model are described in \citet{zha13}.

The mass accretion rate onto the forming protostar $\mdot_*$ is
\begin{eqnarray}
	\mdot_* &=& 9.2 \times 10^{-4}
		\frac{\epsilon_*}{\overline{\epsilon}_*^{0.5}}
		\left(\frac{m_{*}}{M_{\rm c}} \right)^{0.5} \nonumber \\
		&&\times
		\left(\frac{M_{c}}{60\msun} \right)^{3/4}
		\left( \frac{\Sigcl}{1~\gcm}\right)^{3/4}
		\msunyr,
		\label{eq_mdot}
\end{eqnarray}
where $\epsilon_*(t)$ and $\overline{\epsilon}_*(t)$ are the
instantaneous and averaged star formation efficiencies. The
instantaneous star formation efficiency ${\epsilon}_*$ is the ratio
of the accretion rate onto the star to the idealized spherical
infall rate from the initial cloud core in the absence of feedback:
a certain percentage of infalling material is swept up by the disk wind (see next paragraph) and a fraction, $f_d$, of the accreted mass remains in the disk during the protostellar phase.  The averaged star formation efficiency $\overline{\epsilon}_*$ is the ratio of the stellar mass to the idealized collapsed mass of the initial cloud core at the same stage of infall in the absence of feedback. The accretion rate is high for high clump mass surface densities, since high $\Sigcl$ results in a more compact core and thus shorter collapse time.  Also, this formula indicates that in the
no-feedback limit of $\epsilon_*=\overline{\epsilon}_*=1$, the accretion rate increases with time, i.e., as $m_*^{0.5}$ (see MT03).

The magnetically driven bipolar outflows sweep up part of the core and
thus help set the star formation efficiency $\epsilon_*$.  The density
and velocity distributions of the outflow are described by a
semi-analytic disk wind solution that is modified from the
centrifugally driven outflow model of \citet{bla82} \citep[for
    details, see][and ZTH14]{zha13}.  The mass loading rate of the
wind is proportional to the mass accretion rate, with a ratio
$f_w=0.1$ assumed as a typical value \citep{kon00}.  The velocity is
as high as $\sim1000\:{\rm km\:s^{-1}}$ near the rotation axis and
then smoothly decreases at lower latitudes.  This outflow velocity
structure represents features of both a collimated fast jet and a
wider-angle, slower wind.  The opening angle of the outflow
  cavity $\theta_{\rm w, esc}$ is estimated following the method of
\citet{mat00}: if the outflow momentum is strong enough to accelerate
the core material to its escape speed, the outflow extends in that
direction.  The outflow momentum, ${p}_w$, is evaluated
by integrating the momentum rate $\dot{p}_w$ over time. The momentum
injection rate is typically
$\dot{p}_w\simeq (0.15$ -- $3) \times \mdot_*v_{K*}$
in our model, where $v_{K*}\simeq440(m_*/10\msun)^{1/2}(r_*/10\rsun)^{-1/2}\kms$
is the Keplerian speed at the stellar radius.
This process sets the star formation
efficiency $\epsilon_*$ (and also $\overline{\epsilon}_*$).  For
more details of the outflow model and the evaluation of the star formation efficiency,
see \citet{zha13} and ZTH14.

Figure \ref{fig_n} shows an example of the evolution of the density
and velocity structures in the fiducial case.  The gradual opening-up
of the outflow can be seen at the scale of $2\times 10^4\:{\rm~AU}$
with typical outflow velocity of $100\:$--$\:500\:\kms$ which is high
near the axis.  This leads to a decrease in the instantaneous star
formation efficiency.  At the end of accretion in this model, the
stellar mass reaches $26\:\msun$ and the final averaged star formation
efficiency is $\overline{\epsilon}_{*f}=26\msun/60\msun=0.43$.  For
the different models, the averaged final star formation efficiencies
are similar, with a range of
$\overline{\epsilon}_{*f}=0.34\:$--$\:0.52$. While we expect the
momentum flux from protostellar outflows to be the dominant source of
feedback, we note that additional feedback processes due to
ionization, (line-driven) stellar winds and radiation pressure are not
yet included in our modeling, so the above efficiencies may be
somewhat overestimated. The growth of the disk radius can be seen in
the panels showing the scale of $100\:{\rm AU}$.  Note, partly for the
reason of being near the end of its accretion phase and having a
relatively low accretion rate, the disk in the $24\:\msun$ case
appears relatively thin compared to earlier stages (ZTH14).  On the
$100\:{\rm AU}$ scales, we also see the narrow, inner, on-axis cavity
of the disk wind, which is set to have zero (i.e., negligible)
density.  This cavity is too thin to see in the larger scale panels,
but it is resolved by the logarithmic grid spacing.

The evolution of the protostar depends on its accretion history.
ZTH14 calculated the protostellar properties with a detailed stellar
evolution model by \citet{hos09} and \cite{hos10}, self-consistently
adapted to the Turbulent Core Model accretion rates. When the flow
reaches the stellar surface, the accretion energy, $L_{\rm
  acc}=Gm_*\mdot_*/(2r_*)$, is released there.  Following ZTH14, we
treat this accretion luminosity and the internal stellar luminosity as
being emitted isotropically with a single effective temperature,
$T_{\rm *,acc}$, i.e., $L_{\rm *,acc}=L_* + L_{\rm acc}=4\pi r_*^2
\sigma T_{\rm *,acc}^4$.  We will refer to $T_{\rm *,acc}$ as the
stellar effective temperature.  ZTH14 also included the effects of the
accretion energy released at the disk surfaces in the dust temperature
calculation.  However, in this study, we neglect the ionizing photons
from the disk surface, since its surface temperature is not high
enough to emit significant amounts of ionizing flux. If the protostar
rotates at almost its breakup velocity, it distorts the structure of
the star.  This effect leads to nonuniform surface temperatures: the
polar region is hotter than the equatorial region \citep{mey10}.
However, the degree of rotation of massive protostars is quite
uncertain, and thus for simplicity we adopt a spherical model with a
single surface temperature as a first approximation.

Figure \ref{fig_protost} shows the evolution of protostellar
properties with protostellar mass: the stellar radius, luminosity and
effective temperature were calculated by ZTH14, while the evaluation
of ionizing flux will be described in \S \ref{sec_ionlum}.  In the
fiducial case with $M_{c}=60~\msun$ and $\Sigma_{\rm cl}=1\:{\rm
  g~cm^{-2}}$, the stellar radius reaches a maximum at
$m_*\simeq5\:\msun$ when the Kelvin-Helmholz time becomes as short as
the accretion time, after which the protostar is able to begin
contraction towards the zero-age main sequence (ZAMS)
\citep{hos09}. Due to this variation in radius, the effective
temperature increases strongly for $m_*\gtrsim5\msun$.  Comparing to
the fiducial case, the mass at which the stellar radius has a maximum
is larger in the higher $\Sigma_{\rm cl}$ case and smaller in the
lower $\Sigma_{\rm cl}$ case.  This is because a higher $\Sigma_{\rm
  cl}$ causes higher accretion rates (eq. \ref{eq_mdot}), which make
the accretion timescale shorter and also lead to increased swelling of
the protostar.

\subsection{Photoionization}\label{sec_method_photo}

\subsubsection{Ionizing photon luminosity}\label{sec_ionlum}

The total stellar H-ionizing photon luminosity $S_{*,\rm acc}$ is evaluated as
\begin{eqnarray}
	S_{\rm *,acc} = \int_{\nu_0}^{\infty} \frac{L_{\nu,\rm *,acc}}{h\nu} d\nu,
\end{eqnarray}
where $\nu$ is the frequency, $\nu_0$ is the Lyman limit frequency,
and $L_{\nu,{\rm *,acc}}$ is the luminosity per frequency interval
(i.e., the stellar spectrum). The stellar atmosphere model of
\citet{cas04} is adopted to estimate $L_{\nu,{\rm *,acc}}$, which
covers wide ranges of effective temperature and surface gravity.  The
variation of ionizing luminosity is sensitive to the effective
temperature and varies by orders of magnitude (bottom panel of Figure
\ref{fig_protost}).  In the fiducial case of group A, the ionizing
luminosity is as small as $3\times10^{32}\:{\rm~s^{-1}}$ at
$4.7\:\msun$, and quickly increases to $10^{48}\:{\rm~s^{-1}}$ or
larger in the Kelvin-Helmholz contraction phase.  Even at the same
mass of $12\:\msun$, the ionizing luminosity of model Ah12 is five
orders of magnitude smaller than that of model Al12.  Therefore, the
accretion history and the detailed stellar evolution calculation are
essential to study the ionization structure.  We note that the
obtained ionizing photon luminosities are somewhat lower than those in
Figure 2 of \citet{tan14} due to improved calculation of the effects
of line blanketing.

\subsubsection{Transfer of ionizing photons}\label{sec_transfer}

We calculate the extent of the ionized regions using our previously
developed radiative transfer code \citep{tan13}, which allows a
treatment of both direct and diffuse ionizing radiation fields.  We
consider only the ionization of Hydrogen, ignoring Helium and other
elements.  If the stellar effective temperature is high, $T_{\rm
  *,acc}>40,000{\rm K}$, we would underestimate the electron density
by about $10\%$ due to singly ionized Helium, which would then enhance
free-free emission by about $20\%$.

The gray approximation for the photoionization calculation is adopted,
where the frequency of ionizing photons is represented by the mean
value.  The equation of radiative transfer along a ray is
\begin{eqnarray}
	\frac{dI}{ds}
	= - x_{\rm I} n_{\rm H} \sigma_{\rm H} I
		+ \alpha_{1}x_{\rm II}^2n_{\rm H}^2 \frac{\langle h\nu \rangle}{4\pi}  \nonumber \\
	- n_{\rm H} \sigma_{\rm d,a} I 
	+  n_{\rm H} \sigma_{\rm d,s} (J-I),
	\label{eq_RT}
\end{eqnarray}
where $n_{\rm H}$ is the number density of hydrogen nuclei, $I$ is the
irradiance intensity, $J=\int I d\Omega/4\pi$ is the mean intensity,
$\langle h\nu \rangle$ is the mean energy of ionizing photons,
$\sigma_{\rm H}$ is the mean photoionization cross-section of the
Hydrogen atom, $\alpha_{1}$ is the radiative recombination coefficient
to the ground state, $x_{\rm I}$ and $x_{\rm II}$ are neutral (atomic)
and ionized fractions of H, $\sigma_{\rm d,a}$ and $\sigma_{\rm d,s}$
are absorption and scattering cross sections of dust grains per H
nucleon.  The first term of the right-hand side represents the
consumption of ionizing photons by neutral Hydrogen, the second term
describes the re-emission of ionizing photons by recombinations
directly to the ground state, and the third and fourth terms
correspond to absorption and scattering by dust grains, respectively.
The local ionization fraction of H, $x_{\rm II}$, is calculated from
the balance between photoionization, recombination and advection,
\begin{eqnarray}
\nabla \cdot \left( n_{\rm H} x_{\rm II} {\bf v} \right)
= \frac{4\pi x_{\rm I} n_{\rm H} \sigma_{\rm H} J}{\langle h\nu \rangle} - \alpha_{\rm A} x_{\rm II}^2 n_{\rm H}^2, \label{eq_xHII}
\end{eqnarray}
where
${\bf v}$ is the velocity of the disk wind gas,
$\alpha_{\rm A}$ is the recombination coefficient for all levels
(so-called case A).  The advection term on the left-hand side of this
equation can be especially important near the ionization boundary.
The recombination rates, $\alpha_{1}$ and $\alpha_{\rm A}$, are
functions of temperature of the ionized gas.  The mean values of the
photon energy and Hydrogen cross-section are evaluated from the
stellar spectra for each model,
\begin{eqnarray}
	\langle h\nu \rangle &=&
		\frac{1}{S_{*,\rm acc}}
		\int_{\nu_0}^{\infty} L_{\nu,{\rm *,acc}} d\nu, \\
	\sigma_{\rm H} &=&
		\frac{1}{S_{*,\rm acc}}
		\int_{\nu_0}^{\infty} \sigma_{\rm H,\nu}
		\left( \frac{L_{\nu,{\rm *,acc}}}{h\nu} \right) d\nu,
\end{eqnarray}
where $\sigma_{\rm H,\nu}$ is the frequency dependent cross-section of
the Hydrogen atom.  We adopt the fitting formulas from \citet{dra11}
for the recombination rates $\alpha_{1}$ and $\alpha_{A}$, and the
Hydrogen cross-section $\sigma_{\rm H,\nu}$.  The absorption and
scattering properties of dust in these \HII~regions are quite
uncertain.  We adopt $\sigma_{\rm d,a}$ and $\sigma_{\rm d,s}$ equal
to zero in regions where the temperature history has exceeded the dust
destruction temperature (1,400~K) in the thermal radiation calculation
by ZTH14 (i.e., there is a dust-free inner region of the disk wind),
and elsewhere they are fixed at $10^{-21}~{\rm cm^2}$ from a typical
value of the diffuse interstellar medium \citep{wei01}. 
The effect of variation of dust grain properties, i.e., sizes, abundances, compositions, is deferred to a future paper.

We conduct axisymmetric two dimensional calculations to solve
equations (\ref{eq_RT}) and (\ref{eq_xHII}), using our developed
ray-tracing radiative transfer code. In the previous work of
\citet{tan13}, the code solved the transfer equation by the method of
short characteristics, which involves only the cells next to the point
under consideration \citep{sto92}.  Here, we update the code
implementing the method of long characteristics.  With this method,
the transfer problem is solved along the rays from the domain boundary
to each point, which leads to less numerical diffusion and treats the
diffuse radiation more accurately than the short characteristics
method.  The computational domain is a cylindrical region with radial
and vertical coordinates $R\leq20,000\:{\rm~AU}$ and
$Z\leq20,000\:{\rm~AU}$, assuming equatorial plane symmetry.  The
ionizing photon source is a sphere located at $(R,Z)=(0,0)$ with a
radius of $r_{*}$, which represents the accreting star.  We denote the
distance from the center of the star as $r=\sqrt{R^2+Z^2}$.  The
spatial grids for radial and vertical directions are set to be spaced
logarithmically since the inner region has finer structures.  The
entire computational domain is resolved with a $256\times256$ grid.
The radiation intensity $I$ is a function of location and also
direction: $I=I(R,Z,\theta,\phi)$ where $\theta$ and $\phi$ are the
{\rm polar} and azimuthal angles, respectively.  The radiation emitted
from the vicinity of the star has very fine angular structure when it
reaches the outer boundary, i.e., $\delta \theta \sim \delta \phi \sim
r_*/r_{\rm max}\sim 10^{-6}$.  To resolve these fine structures while
still saving computational costs, the angular resolutions of $\theta$
and $\phi$ are configured to vary depending on location $(R,\;Z)$.  In
the $\theta$ direction, the radiation arriving from directions with
$\theta \simeq \tan^{-1}(R/Z)$, which comes from the vicinity of the
star, is resolved with a high resolution of $\Delta \theta \ll r_*/r$,
while the radiation arriving from $\theta \simeq \tan^{-1}(R/Z)+\pi/2$
is resolved by $\Delta \theta\simeq10^{-2}$.  In the $\phi$ direction,
based on the tangential plane method \citep{sto92}, the radiation
emitted from near the axis and from the opposite direction,
$\phi\simeq0$ and $\pi$, is resolved with high resolution of $\Delta
\phi \ll r_*/r$, while the radiation from $\phi\simeq\pi/2$ is
resolved by $\Delta \phi \simeq 10^{-2}$.  We also conduct resolution
tests for the fiducial case with grids of $128\times128$ and
$330\times330$.  The obtained ionized structures are consistent at
these resolutions and free-free fluxes change by less than $6\%$ at
radio frequencies from $0.3\:$--$\:300\rm GHz$.

In this first work we study the photoionization of the disk wind, and
do not allow photoionization of gas in the disk or infall envelope
components of the model. Such photoionization would enhance the mass
loss rate that is being injected into the disk wind, i.e., by
photoevaporation, and would thus change its density structure and also
the star formation efficiency.  This feedback of photoevaporation will
be discussed in a future paper.

\subsubsection{Temperature of the ionized gas} \label{sec_T}

The temperature of the ionized gas is important to evaluate
recombination rates and to study free-free emission for observational
diagnostics.  The photoionized gas temperature is regulated to around
$10,000{~\rm K}$ by the balance of heating and cooling processes.  The
dominant heating process is photoionization and the main cooling
processes are radiative cooling of collisionally excited metal lines,
recombination of H, and free-free emission.

It is computationally expensive to take into account all heating and
cooling processes at every step of the radiative transfer calculation,
and thus here we introduce an approximate treatment.  The heating rate
by photoionization depends on the stellar spectrum.  Also, heating and
cooling rates are both mostly sensitive to the density.  Therefore, we
treat the photoionized gas temperature, for each stellar model, as a
function of the local density, i.e., $T(n_{\rm H})$.  We evaluate
$T(n_{\rm H})$ from the volume average temperature of static
\HII~regions with uniform densities in the range $n_{\rm
  H}=10\:$--$\:10^{11}\:{\rm cm^{-3}}$ using version 13.03 of
\cloudy~\citep{fer13}.  This approximate approach is appropriate since
a uniform density \HII~region for each stellar model is almost
isothermal as shown below. It should be noted that the ionized gas
temperature is also sensitive to the metallicity of the gas, and we
assume solar metallicity in this study.

\begin{figure}
\begin{center}
\plotone{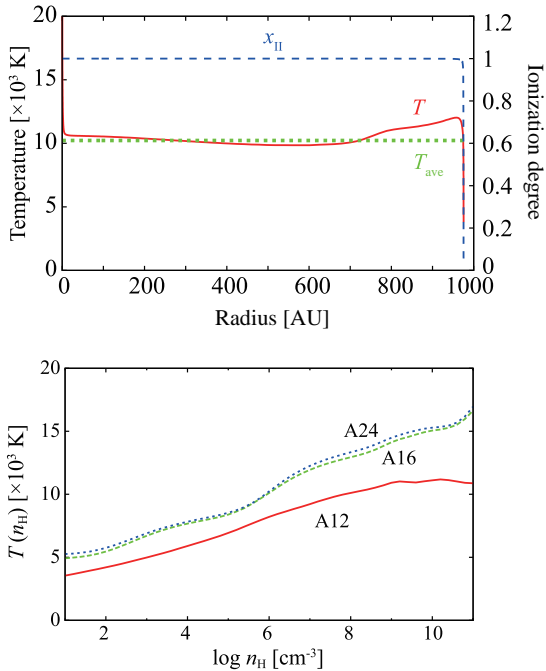}
\end{center}
\caption{
{\it Top}: Radial structure of an \HII~region with an uniform density
of $n_{\rm H}=10^6\:{\rm cm}^{-3}$ formed by the radiation of the
stellar model of A24.  The radial structures of temperature, $T$,
(solid) and ionization degree of H, $x_{\rm II}$, (dashed) are shown.
The volume averaged temperature of $T_{\rm ave}=10,225{\rm~K}$ is also
indicated (dotted).  {\it Bottom}: Ionized gas temperature as a
function of density, $T(n_{\rm H})$, which is set equal to the
volume-averaged photoionized gas temperature, $T_{\rm ave}(n_{\rm
  H})$. Results are shown for stellar models A12, A16, and A24.}
\label{fig_uniform}
\end{figure}

The top panel of Figure \ref{fig_uniform} shows the radial structure
of an \HII~region with uniform density of $n_{\rm
  H}=10^6\:{\rm~cm^{-3}}$ formed by the stellar radiation of the
stellar model A24.  It can be seen that the \HII~region has a sharp
edge at $960\:{\rm AU}$ where the ionization degree drops from near
unity to near zero. The temperature is almost isothermal in the
ionized region.  We evaluate the typical photoionized gas temperature
at $n_{\rm H}=10^{6}\:{\rm cm^{-3}}$ as being $10,225 {\rm K}$ from
the volume average of this temperature for this stellar model.

In this way, we obtain photoionized gas temperatures over the density
range of $n_{\rm H}=10\:$--$\:10^{11}\:{\rm cm^{-3}}$ for each stellar
model.  The bottom panel of Figure \ref{fig_uniform} shows the
photoionized gas temperature as functions of density, $T(n_{\rm
  H})=T_{\rm ave}(n_{\rm H})$, for the stellar models A12, A16, and
A24.  The temperatures of all models are around $5,000$--$15,000 {\rm
  K}$ for orders of magnitude of density variation.  The temperature
is typically higher at higher density.  This is because the cooling is
suppressed if the density exceeds the critical density of some of the
cooling lines.  We adopt these averaged temperatures to set $T(n_{\rm
  H})$ as the local temperature of the ionized gas in our radiative
transfer calculations.

Since the adopted temperature is evaluated based on static \HII~region
calculations, adiabatic cooling has not been included in this study.
The adiabatic cooling rate, $\Lambda_{\rm ad}\simeq 2.5\nabla \cdot
(n_{\rm H} k_{\rm B} T{\bf v})$, is proportional to $\sim n_{\rm
  H}/r$.  On the other hand, the radiative cooling rates in ionized
regions are almost proportional to the square of density.  Since the
density gradient in the disk wind is as steep as $d \log n_{\rm H}/d
\log r \sim -1.5$ to $-2$, adiabatic cooling may not be negligible in
low density regions far from the star.  We compare the adiabatic
cooling rate to the cooling rate evaluated by the static calculations.
In the fiducial model of A16, adiabatic cooling would be $>20\%$ of
total cooling rate when $n_{\rm H}<10^4\:{\rm cm^{-3}}$ and
$r>1000\:{\rm AU}$.  As we will see in \S\ref{sec_SED}, the free-free
flux at frequencies lower than $1{\:\rm GHz}$ is mainly provided from
$r>1000{\rm AU}$.  Therefore, we note that the SEDs obtained by our
simplified modeling would be overestimated somewhat at $\la1\:{\rm
  GHz}$.

\subsection{Observables}

\subsubsection{Free-free emission}

Using the obtained structures of the photoionized outflow, we estimate
the free-free radio continuum emission. The equation of transfer for
free-free emission is
\begin{eqnarray}
	\frac{dI_{\nu,{\rm ff}}}{ds} = -\kappa_{\nu,{\rm ff}} \left(I_{\nu,{\rm ff}}  - B_\nu \right),
	\label{eq_RTff}
\end{eqnarray}
where $I_{\nu,{\rm ff}}$ and $\kappa_{\nu,{\rm
    ff}}$ are the intensity and opacity of free-free emission, and
$B_\nu(T)$ is the Planck function.  The opacity due to free-free
absorption is given by
\begin{eqnarray}
\kappa_{\nu,{\rm ff}} = \frac{4}{3} \left( \frac{2\pi}{3} \right)^{1/2}
		\frac{e^6}{m_{e}^{3/2}hc} \frac{n_{\rm H}^2 x_{\rm HII}^2}{\sqrt{k_{B}T}} \nonumber \\ 
		\times
		\left[ \frac{1-\exp{\left( -h\nu/k_{B}T \right)} }{\nu^3} \right]
		g_{\rm ff}, \label{eq_kapff}
\end{eqnarray}
where $e$ and $m_e$ are the charge and mass of the electron, $h$ is
the Planck constant, $c$ is the speed of light, $k_{B}$ is the
Boltzmann constant, and $g_{\rm ff}(\nu,T)$ is the Gaunt factor for
free-free emission. The observed flux density is obtained by
integrating over source solid angle of $\Omega_s$,
\begin{eqnarray}
	F_{\nu,{\rm ff}}
	= \int_{\rm \Omega_s} I_{\nu,{\rm ff}} \cos(\theta-\theta_{\rm view})  d\Omega
	= \int_{\rm \Omega_s} I_{\nu,{\rm ff}} d\Omega, \label{eq_fff}
\end{eqnarray}
where $\theta_{\rm view}$ is the viewing inclination measured from the
outflow axis.  In the last formula, we assumed that the distance is
much farther than the object size, i.e., $\theta - \theta_{\rm view}
\la 20,000{\rm AU}/1{\rm\:kpc} \ll 1$.  In this study, for
quantitative results, we assume protostars are located at a distance
of $1\:{\rm~kpc}$, and total fluxes are obtained by integrating over a
radius out to $10^4\:{\rm~AU}$ from the central star (for all our
models this region dominates the total flux from free-free emission).
The free-free emission is typically observed at
$\nu\simeq0.1$--$100\:{\rm~GHz}$, while the thermal dust emission
would usually be dominant at higher frequencies.  However, for
completeness, we calculate free-free emission over the frequency range
of $\nu=10^{-1}$--$10^7{\rm~GHz}$.

\subsubsection{Hydrogen recombination lines}

We also evaluate the HRL emission from the photoionized outflow.
Considering a H$n\alpha$ line, i.e., the line from the transition $n+1
\rightarrow n$, the opacity is
\begin{eqnarray}
	\kappa_{\nu, \rm HRL} &=& n_n \frac{\lambda_{n}^2}{8\pi}
	\left( \frac{n+1}{n} \right)^2 \nonumber \\
	&&\times A_{n+1,n} \left( 1-e^{-h\nu_n/k_{B}T} \right)
	\phi_\nu, \label{eq_kapl}
\end{eqnarray}
where $\phi_\nu$ is the normalized line profile ($\int \phi_\nu
d\nu=1$), $n_n$ is the population number in energy level $n$,
$A_{n+1,n}$ is the Einstein coefficient of spontaneous emission, and
$h\nu_{n}$ and $\lambda_{n}=c/\nu_{n}$ are the energy and wavelength
of the H$n\alpha$ transition, respectively.  Here, for simplicity, we
assume the case of local thermodynamic equilibrium (LTE).  However we
note that, in the particular case of the massive star MWC 349A, maser
amplification of HRLs are observed and thus non-LTE effects can be
important \citep[for modeling of MWC 349A including non-LTE effects, see]
[]{bae13}.

The transfer for the total intensity, $I_{\nu, \rm ff, HRL}$, i.e.,
the sum of the free-free and HRL emission, is
\begin{eqnarray}
	\frac{dI_{\nu,{\rm ff,HRL}}}{ds} &=& -(\kappa_{\nu,{\rm ff}}+\kappa_{\nu,{\rm HRL}}) \left(I_{\nu,{\rm ff,HRL}} - B_\nu \right),
\end{eqnarray}
and the total observed flux is
\begin{eqnarray}
	F_{\nu,{\rm ff,HRL}}
	&=& \int_{\rm \Omega_s} I_{\nu,{\rm ff,HRL}}
		\cos(\theta-\theta_{\rm view})d\Omega \nonumber \\
	&=& \int_{\rm \Omega_s} I_{\nu,{\rm ff,HRL}} d\Omega.
\end{eqnarray}
Here, we assumed that the source distance is greater than its size as
in equation (\ref{eq_fff}).  We evaluate the HRL flux as the excess
from the continuum flux at $\nu_n$, i.e., $F_{\nu,{\rm
    HRL}}=F_{\nu,{\rm ff,HRL}}-F_{\nu_n,{\rm ff}}$.  Of course,
observationally it is not possible to observe the free-free flux
exactly at $\nu_n$ since the line exists there and we cannot divide
the total flux into two components.  However, the value of
$F_{\nu_n,{\rm ff}}$ is still able to evaluated from the far tail of
the line profile or from interpolating from continuum observations at
other frequencies.  Since the typical line width is $\Delta \nu/\nu_n
\sim v/c< 10^{-3}$ and
the typical free-free emission spectral index in this study
is $0\la p\la1$, the variation of the continuum flux in this width is less than
$0.1\%$.  Thus, it is safe to assume the continuum is constant over
the line width.  In this paper we will illustrate these results with
the H$40\alpha$ line at $99\:{\rm GHz}$.  The flux ratio is typically
$F_{\nu,{\rm HRL}}/F_{\nu,{\rm ff}}\sim0.1$ for the H$40\alpha$ line
in our models.

We assume that locally the gas has a Maxwellian velocity distribution,
ignoring the intrinsic width of the line relative to the thermal and
bulk-motion broadening.  The velocity dispersion of Hydrogen is
$\sigma_{v}=\sqrt{k_{\rm B}T/m_{\rm H}}$, and the corresponding
dispersion in frequency is $\sigma_{\nu}=(\sigma_{v}/c)\nu_{n}$.  The
local normalized line profile $\phi_\nu$ combining the bulk gas
velocity along the line of sight, $v$, is
\begin{eqnarray}
	\phi_\nu=\frac{1}{\sqrt{2\pi\sigma_{\nu}^2}}
		\exp\left[ -\frac{ \left\{ (1+v/c)\nu-\nu_n \right\}^2 }{2\sigma_{\nu}^2} \right].
		\label{eq_lineprofile}
\end{eqnarray}
The frequency $\nu$ observationally represents a radial velocity of
$(1-\nu/\nu_n)c$.

\subsection{Escape fraction of ionizing photons}\label{sec_esc_model}

As we will see in next section, the ionized region, after initial
early expansion, becomes unconfined in the radial direction along the
outflow axis.  Thus, some fraction of the ionizing photons escapes
from the protostellar core and can ionize the surrounding, ambient
clump gas, which may be forming a stellar cluster.  Although the main
focus of this work is the ionized structure of disk winds and their
observational properties, ambient gas ionization is also important for
its effect on star cluster formation and production of additional
radio emission.  Therefore, we provide estimates of the ionization of
escaping photons and radio emission from the ionized clump gas.

The escape fraction of ionizing photons depends on the polar angle
$\theta$, i.e., in response to the ionized disk wind structure.  We
calculate the escape fraction as
\begin{eqnarray}
	f_{\rm esc}(\theta) = \frac{4\pi r^2 F_{r,\rm out}(\theta)}{S_{\rm *,acc}\langle h\nu \rangle},
\end{eqnarray}
where $F_{r, \rm out}$ is the radial energy flux of ionizing photons
at the outer boundary, i.e., $R=20,000\:{\rm AU}$ or $Z=20,000\:{\rm
  AU}$. We note that the escape fraction is insensitive to the choice
of the radius at which escape is defined as long as it is
significantly larger than the disk radius.  Integrating $f_{\rm esc}$
over the boundary, we obtain the total escape fraction,
\begin{eqnarray}
	\overline{f}_{\rm esc} = \frac{1}{4\pi}\int f_{\rm esc}(\theta) d\Omega.
\end{eqnarray}

\section{Results}\label{sec_results}

\subsection{Structure and evolution of the photoionized outflow}\label{sec_outflow}

\begin{figure}
\begin{center}
\includegraphics[width=90mm]{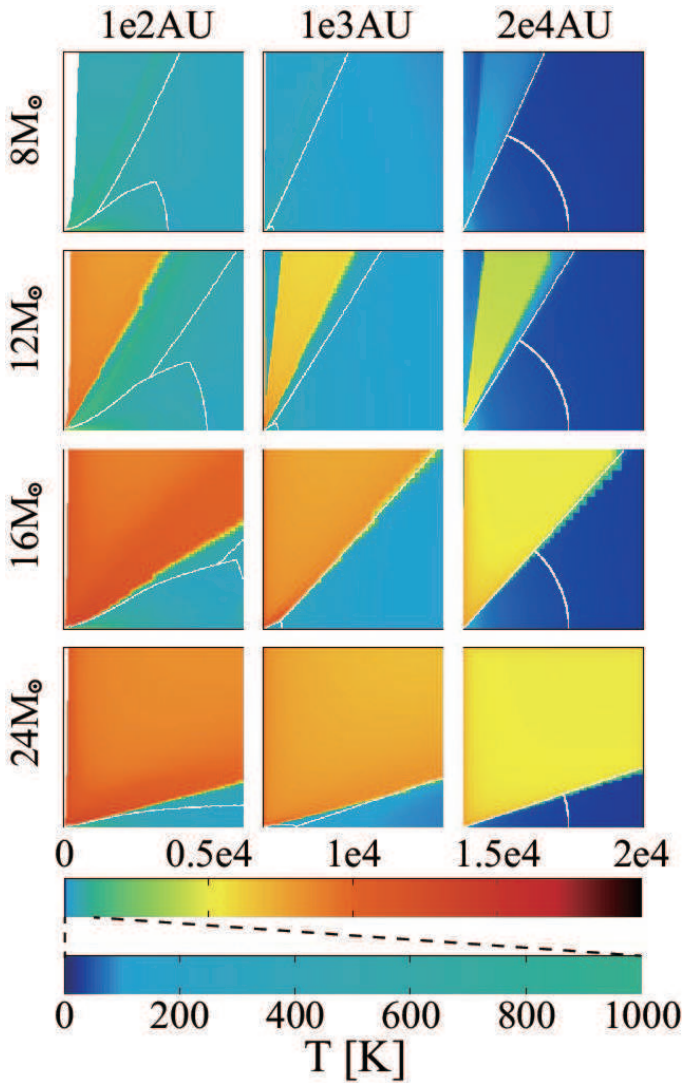}
\end{center}
\caption{
Evolutionary sequences of the photoionized outflows of model group
A. The color scale shows the temperature of the photoionized and
neutral regions.  The white regions seen close to the vertical axes in
the $100{\rm\:AU}$-scale panels (left column) are disk wind inner
cavities, where the gas density is negligible (formally zero) and the
ionizing photon can travel freely.  As in Figure \ref{fig_n}, the core
scale and the boundaries between disk, core envelope, outflow and
clump are shown by white lines.}
\label{fig_T_a}
\end{figure}

\begin{figure}
\begin{center}
\includegraphics[width=90mm]{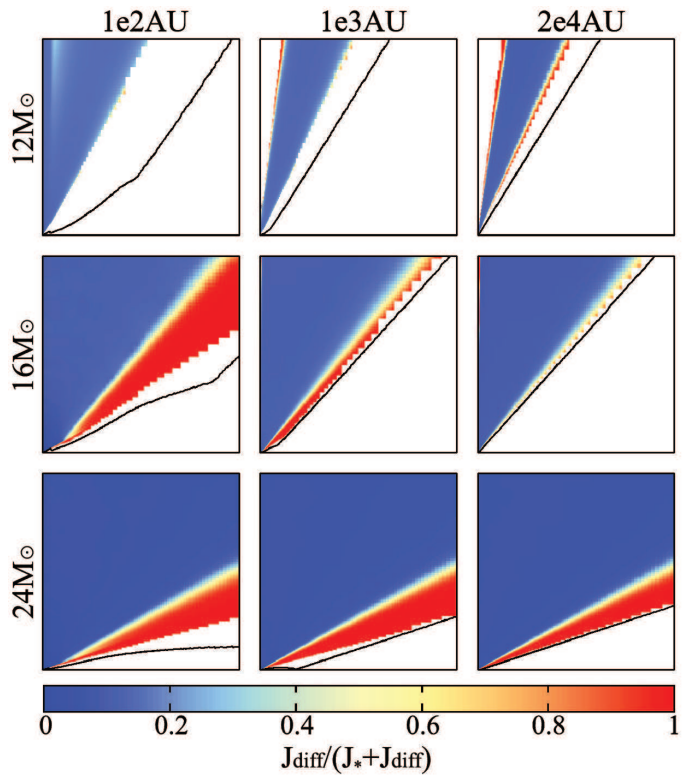}
\end{center}
\caption{
The radiation fields for model group A.  The color scale shows the
fraction of diffuse radiation in the ionized region (the white region
is neutral). Black lines indicate the disk/envelope boundaries with
the disk wind.}
\label{fig_diff}
\end{figure}

\begin{figure}
\begin{center}
\includegraphics[width=75mm]{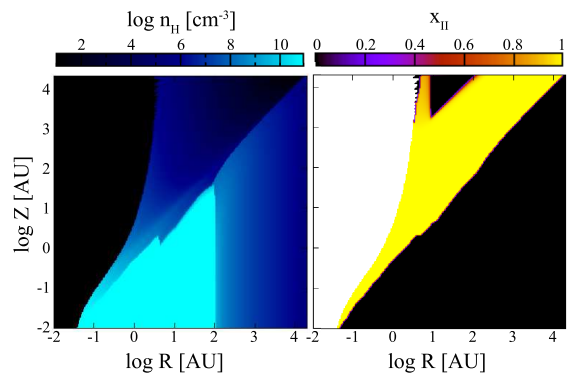}
\end{center}
\caption{
The structures of density $n_{\rm H}$ (left) and ionization degree
$x_{\rm II}$ (right) for model A16 in {\it logarithmic} scale.
The white region in the right panel is the inner hole where the ionization degree
cannot be defined due to the zero density.  However, the ionizing photons
travels this inner hole freely.  Note that the stellar radius is
$r_*=6.4\rsun=0.03{\:\rm AU}$ for this model.  }
\label{fig_log_A16}
\end{figure}

First we report results of the fiducial case, model group A.  Figure
\ref{fig_T_a} shows the evolution of the photoionized structure.  The
high temperature region with $\sim10^4\:{\rm K}$ is the photoionized
gas, i.e., the \HII~region, and elsewhere is neutral gas (with
temperatures calculated from the ZTH14 radiative transfer models;
note, we expect there would be some localized changes to the ZTH14
results for the neutral gas due to the presence of the \HII~region and
its associated photodissociation region (PDR), but we ignore those
here---the PDR region will be considered in a future paper). Note that
the white regions seen along the rotation axes in the $100{\:\rm
  AU}$-scale panels are disk wind inner cavities, where the density is
negligible (formally zero in these models since we do not include a
line-driven wind from the stellar surface) and thus where ionizing
photons can travel freely.  The ionized gas temperature is slightly
higher in the inner region of the disk wind due to the higher density
(see Fig. \ref{fig_uniform}).  When $m_*\leq8\msun$, the ionizing
photon luminosity is too low to ionize the disk wind. As the ionizing
luminosity increases with mass, the photoionized region is formed.
When $m_*=12\:\msun$, the photoionized region is confined to
intermediate zenith angles of $10\degr\la\theta\la30\degr$.
This is because densities are higher near the outflow axis at $\theta<10\degr$
and near its launching zone for $\theta\ga30\degr$.
Since the radial density gradient in the outflow is typically steeper than $r^{-1.5}$,
the photoionized region is not confined in the radial direction and
breaks out. The opening angle of the photoionized region increases as
the protostellar mass grows. The ionized region eventually reaches the
flared disk surfaces, however some parts of the disk wind remain
neutral due to the shielding by the inner disk and disk wind even
though the diffuse radiation is considered. Therefore,
photoevaporation from the disk surface is suppressed, while
photoevaporation from the envelope may still be efficient.

The ionizing radiation has two components: one is the direct stellar
radiation and the other is the diffuse radiation from recombinations
to the ground state and dust scattering.  Figure \ref{fig_diff} shows
the fraction of diffuse radiation, $J_{\rm diff}/(J_*+J_{\rm diff})$,
at each evolutionary stage.  The direct radiation dominates over most
of the outflow.  However, in some locations, due to the direct
radiation being shielded by the inner disk at $\ll 10^2\:{\rm AU}$,
the diffuse radiation becomes more important, i.e., near the
ionized-neutral boundary, e.g., as seen in model A24.  In this way, as
suggested by \citet{hol94}, the diffuse radiation is important to
determine the structure of the \HII~region and rates of
photoevaporation from the neutral components. For these reasons,
accurate treatment of the radiative transfer including the diffuse
component is necessary.

\begin{figure*}
\begin{center}
\includegraphics[width=150mm]{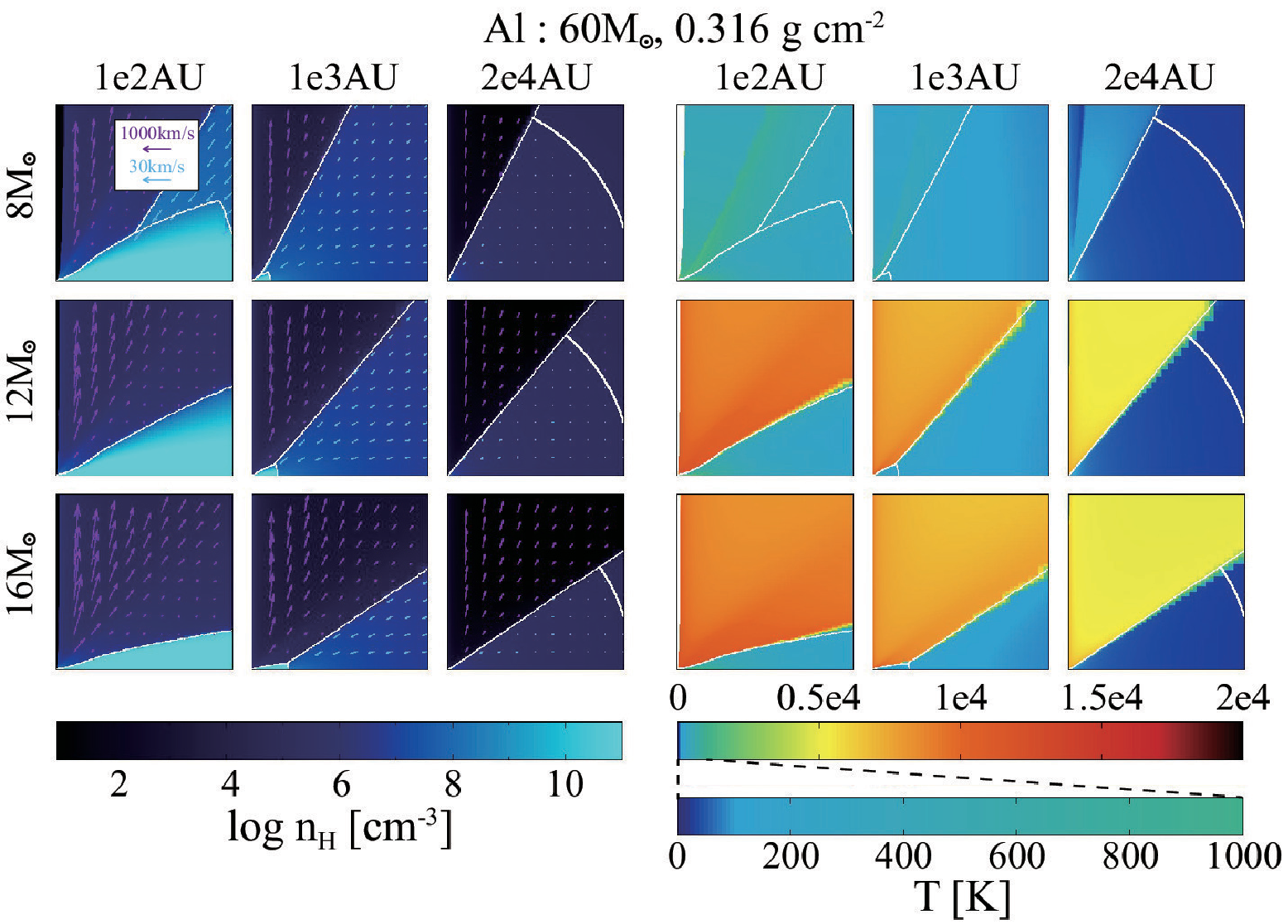}
\end{center}
\caption{
The evolution of density (left) and photoionization (as revealed by gas
temperature, $T$) (right) structures for model groups Al
($M_c=60\:\msun,\:\Sigcl=0.316\gcm$).}
\label{fig_al}
\end{figure*}

\begin{figure*}
\begin{center}
\includegraphics[width=150mm]{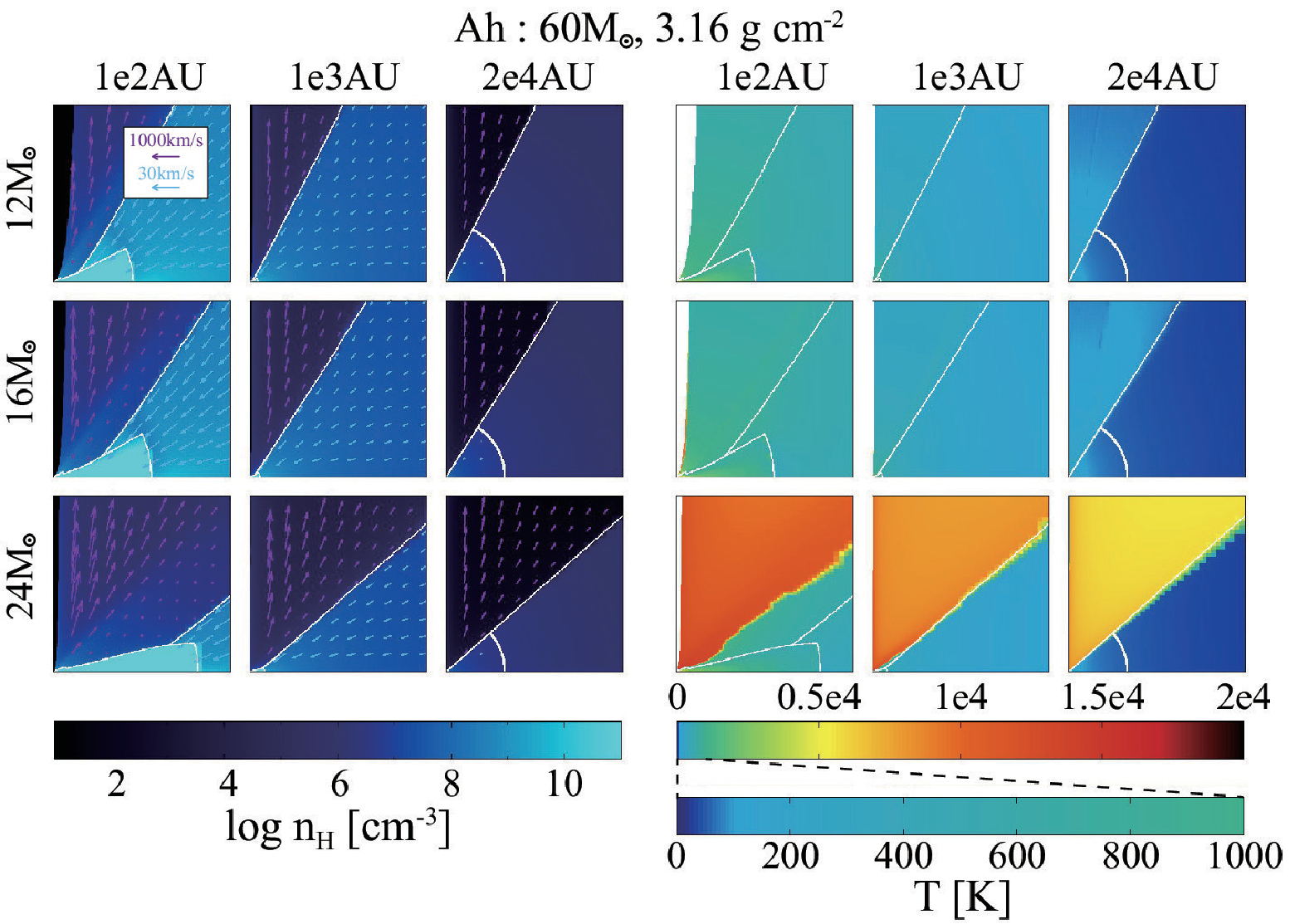}
\end{center}
\caption{
The evolution of density (left) and photoionization (as revealed by gas
temperature, $T$) (right) structures for model groups Ah
($M_c=60\:\msun,\:\Sigcl=3.16\gcm$).}
\label{fig_ah}
\end{figure*}

\begin{figure*}
\begin{center}
\includegraphics[width=150mm]{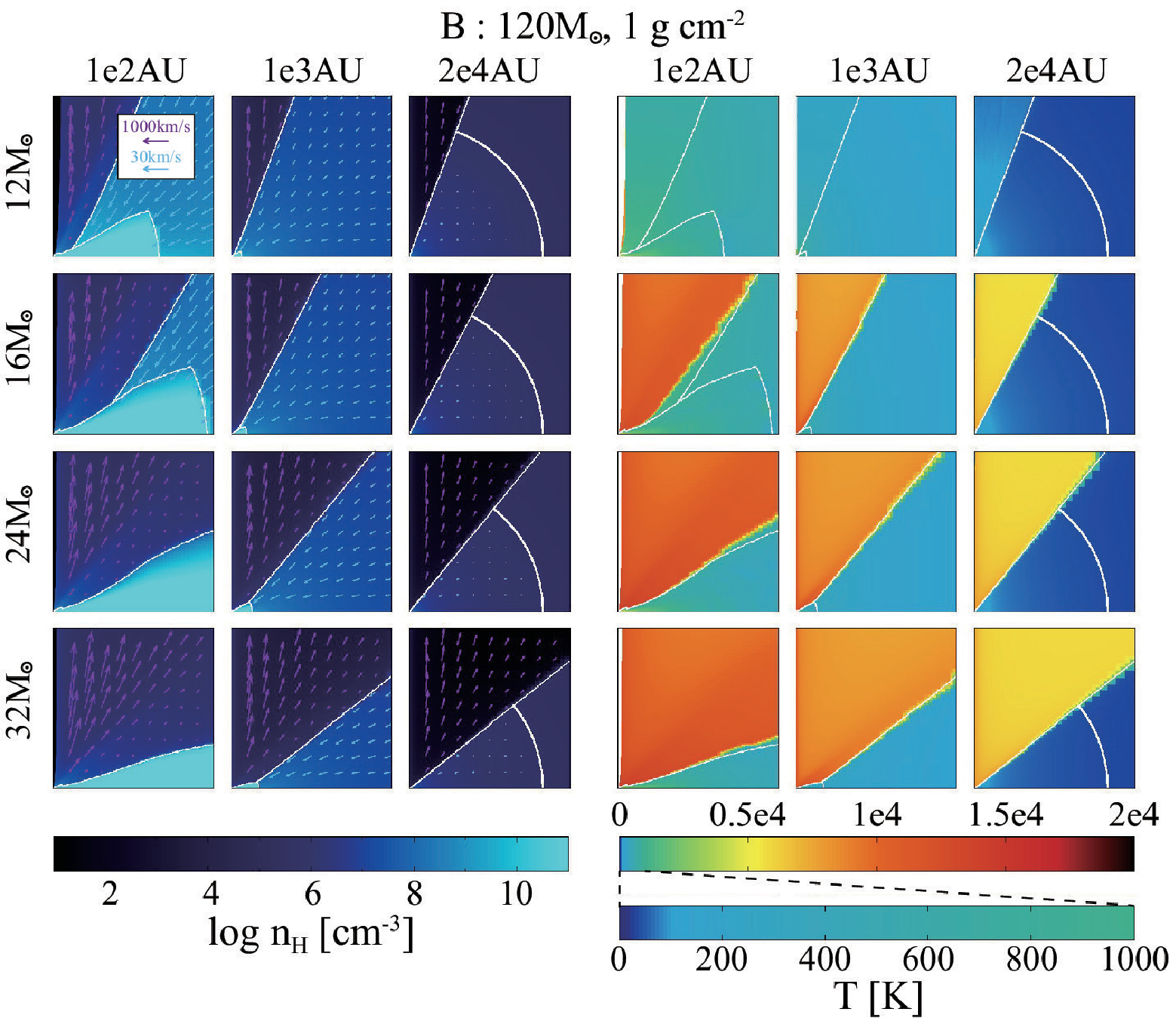}
\end{center}
\caption{
The evolution of density (left) and photoionization (as revealed by gas
temperature, $T$) (right) structures for model groups B
($M_c=120\:\msun,\:\Sigcl=1\gcm$).}
\label{fig_b}
\end{figure*}

\begin{figure*}
\begin{center}
\includegraphics[width=150mm]{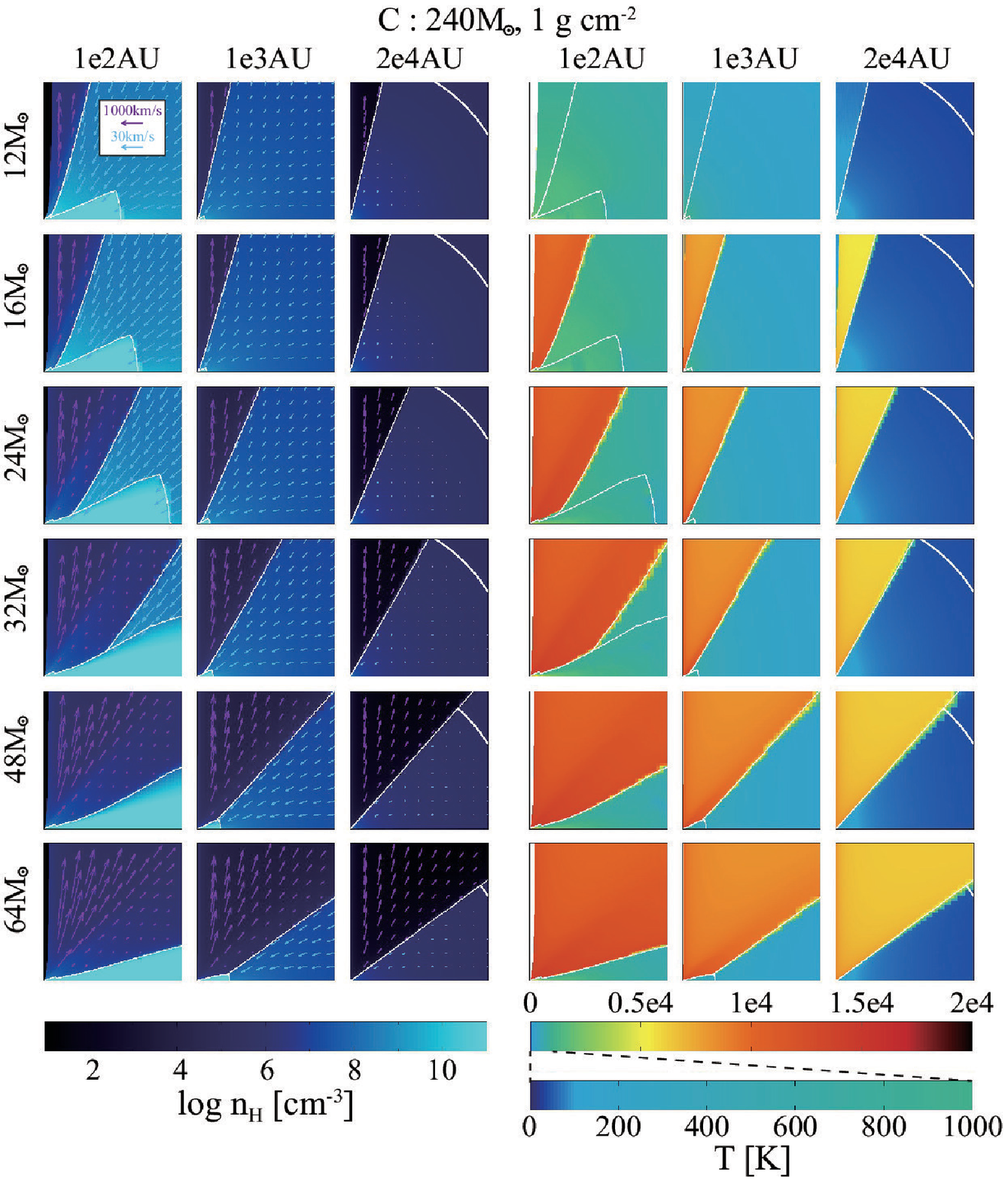}
\end{center}
\caption{
The evolution of density (left) and photoionization (as revealed by gas
temperature, $T$) (right) structures for model groups C
($M_c=240\:\msun,\:\Sigcl=1\gcm$).}
\label{fig_c}
\end{figure*}

To illustrate the structure near the rotation axis,
Figure~\ref{fig_log_A16} shows the density and ionization fraction of
model A16 with a logarithmic scale.  The disk wind inner cavity, with
zero density inside its innermost streamline, is seen as a very
collimated structure with $\theta<2\times10^{-4}$ at $Z=20,000\:{\rm
  AU}$ (the black region in left panel and the white region in right
panel). The density is high along the inner edge of the outflow.  Due
to the shielding by this high density inner wall, the gas in the
region at high latitude with 
$R\sim10$--$100 {\rm\:AU}$ and $Z\ga1000{\rm\:AU}$ remains neutral.
This shielding is the same mechanism seen in model A12 in which the
neutral region is at $\theta\la10\degr$.  
The strip of ionized gas entering this neutral region is caused by
high velocity advection near the axis (eq. \ref{eq_xHII}).  These thin
structures near the axis are sensitive to our adoption of an idealized
axisymmetric model.  In reality, the outflow winds are not expected to
be perfectly axisymmetric. For example, the jet in IRAS 16547-4247
appears to undergo precession \citep{rod08}.  However, the idealized
structures in our model illustrate the wind shielding effect and also
demonstrate the accuracy of our numerical calculation even at such
fine angular scales.

Next we show results with different initial cores with lower and higher $\Sigma_{\rm cl}$.
The evolution of density and photoionization (as revealed by gas temperature, $T$) structures in groups Al and Ah
are shown in Figures~\ref{fig_al} and \ref{fig_ah}.
In group Al, the photoionized region
breaks out when $8\:\msun<m_*<12\:\msun$, as in the fiducial case.
On the other hand in group Ah, the photoionized region is tightly
confined to a thin layer at the inner wind wall even at $16\:\msun$ and only expands later.
This is mostly because higher $\Sigma_{\rm cl}$ leads to higher accretion rates (eq. \ref{eq_mdot}),
which result in a higher mass at which the ionizing luminosity shows its
dramatic increase due to Kelvin-Helmholz contraction (Fig.~\ref{fig_protost}).
We discuss the breakout of outflow-confined HII regions in \S\ref{sec_confined}.

Figures \ref{fig_b} an \ref{fig_c} show the evolution of density and
photoionization (as revealed by gas temperature, $T$) structures in higher $M_{c}$ models of groups B and C.
In both cases, the photoionized region breaks out when $m_*>12\:\msun$.
This is similar to the high $\Sigma_{\rm cl}$ case,
as the higher-mass core leads to delayed (in $m_*$) contraction via higher accretion rates.  If we
ignore feedback effects, the accretion rate is proportional to
$M_{c}^{0.25}$ at the same stellar mass (eq. \ref{eq_mdot}).
In this way, the mass at which the photoionized region emerges varies from
$\sim10\:$--$\:15\:\msun$ depending on its parent cloud core.

\subsection{Free-free continuum emission}

We conduct the radiative transfer calculation of free-free emission
from the obtained photoionized structures.  The total flux, spectrum,
and size of the free-free emission of our model are similar to those
of the radio winds and jets associating with massive star forming
regions as we will see in \S\ref{sec_comparison}. However, our ionized
region model has a limited cylindrical volume (\S\ref{sec_transfer})
and the contribution from external ionized regions is not
included. Thus, it should be kept in mind that the free-free flux
would be enhanced by the emission from the ionized clump if it is on
the line of sight, i.e., $\theta_{\rm view} < \theta_{\rm w, esc}$.

\begin{figure*}
\begin{center}
\includegraphics[width=150mm]{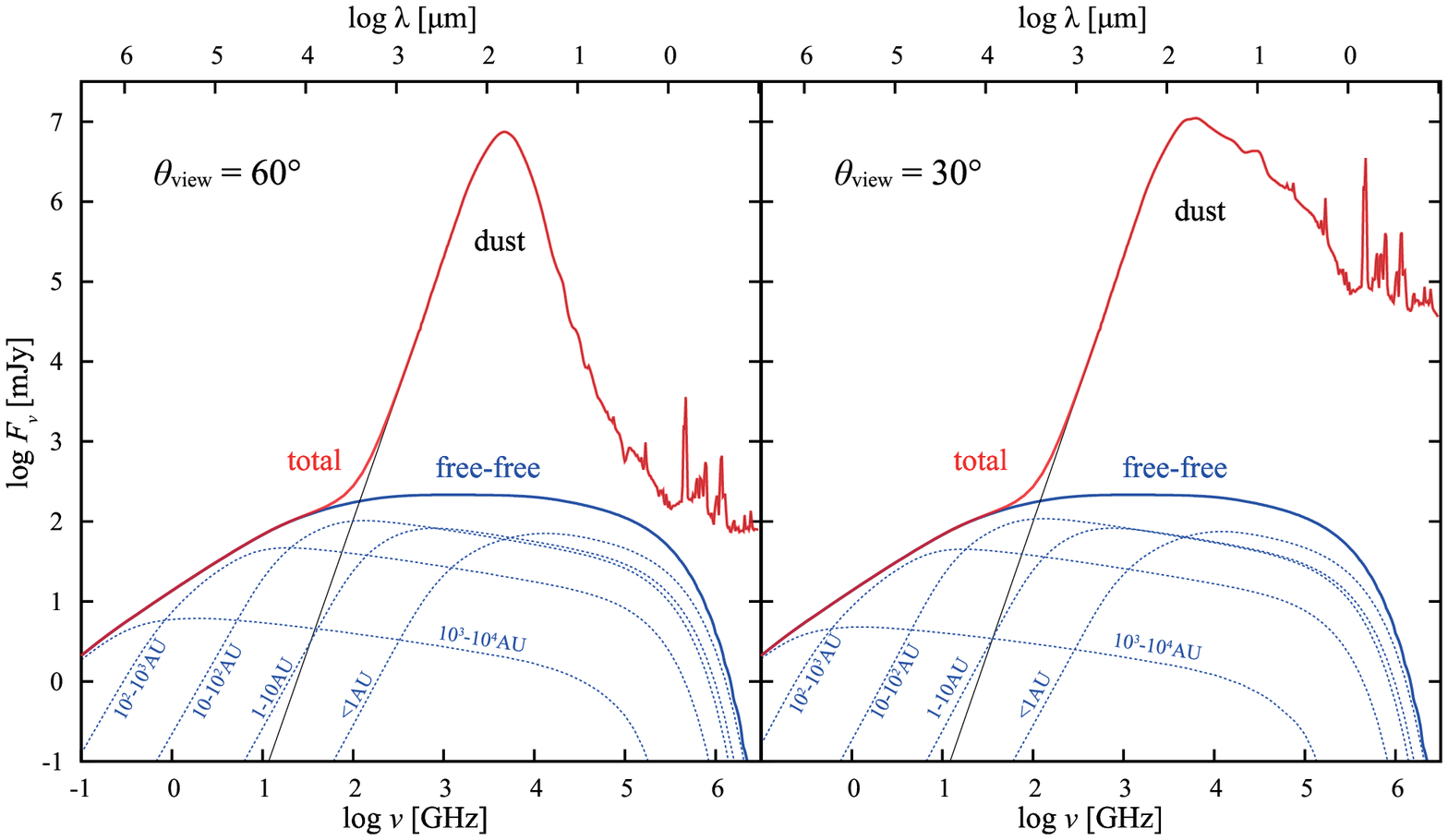}
\end{center}
\caption{
SEDs for model A16 ($M_{c}=60\:\msun$, $\Sigcl=1\:\gcm$, and
$m_*=16\:\msun$) at inclination angles of $60\degr$ (left) and $30\degr$
(right).  Two emission components, free-free (thick blue) and thermal
dust (thin black), are shown in each panel. The blue dotted
lines show the contributions by the regions at $1,10,10^2,10^3$ and
$10^4\:{\rm AU}$ from the central star.  A distance of $1\:{\rm kpc}$ is
assumed for the flux normalizations.}
\label{fig_SEDa16}
\end{figure*}

\subsubsection{Spectral Energy Distributions}\label{sec_SED}

For all SED calculations, a distance of $1\:{\rm kpc}$ and an aperture
diameter of $20\arcsec$ (20,000~AU) are assumed.
Figure~\ref{fig_SEDa16} shows full SEDs including thermal dust and
free-free emission for model A16 at viewing inclinations of $60\degr$
and $30\degr$ of the line of sight to the rotation/outflow axis.  The
flux from thermal dust emission was calculated by ZTH14.  The
free-free emission dominates over the thermal dust emission at radio
frequencies of $\la100\:{\rm GHz}$, while the dust flux dominates at
higher frequencies.  We note that, at wavelengths shorter than
$100\:{\rm \micron}$, we do not count the free-free emission in the
total flux since we do not treat the dust extinction in the free-free
radiative transfer calculation (eq. \ref{eq_RTff}).

The infrared (IR) dust flux (at relatively short wavelengths) depends
strongly on viewing angle.  This is because the stellar radiation is
absorbed by the dense disk and core envelope along directions close to
the equatorial plane, and the re-emitted IR radiation, still suffering
from high optical depths, escapes along the polar, low-density outflow
cavity. Therefore, the observed flux is higher if the line of sight is
passing through the outflow cavity (ZTH14). This process is known as
the ``flashlight effect,'' which is also important for helping to
overcome radiation pressure feedback
\citep{nak95,yor99,kru05,kru09,kui10,kui11}.

On the other hand, the radio flux is almost independent of viewing
angle since it typically experiences much smaller optical depths than
the IR radiation.  We can approximate the spectrum of free-free
emission as being a power law in frequency,
$F_{\nu,{\rm ff}}\propto \nu^{p}$, with spectral index $p$.
From $3$ -- $30\:{\rm GHz}$, the
free-free flux from model A16 is fitted as
$F_{\nu,{\rm ff}}=81(\nu/{\rm GHz})^{0.48}\:{\rm mJy}$.
The spectral index of the free-free emission
decreases from $p=0.9$ at 0.1~GHz to $0.2$ at 100~GHz.
An index smaller than two indicates that the photoionized disk wind is only
partially optically thick with respect to its free-free emission.
The overall spectral index (including thermal dust emission) has a minimum
of $0.45$ at $35\:{\rm GHz}$, and quickly increases at higher
frequencies approaching the dust spectrum value of $p=3.3$.
Figure~\ref{fig_SEDa16} also indicates that the free-free flux at
$\sim 0.1$~GHz is mostly emitted from $\sim10^3$--$10^4$~AU scales,
while that at $\sim10$~GHz is mostly probing $\sim10^2$--$10^3$~AU
scales.  Thus, the higher frequencies trace the structure of the inner
wind.

\begin{figure*}
\begin{center}
\includegraphics[width=120mm]{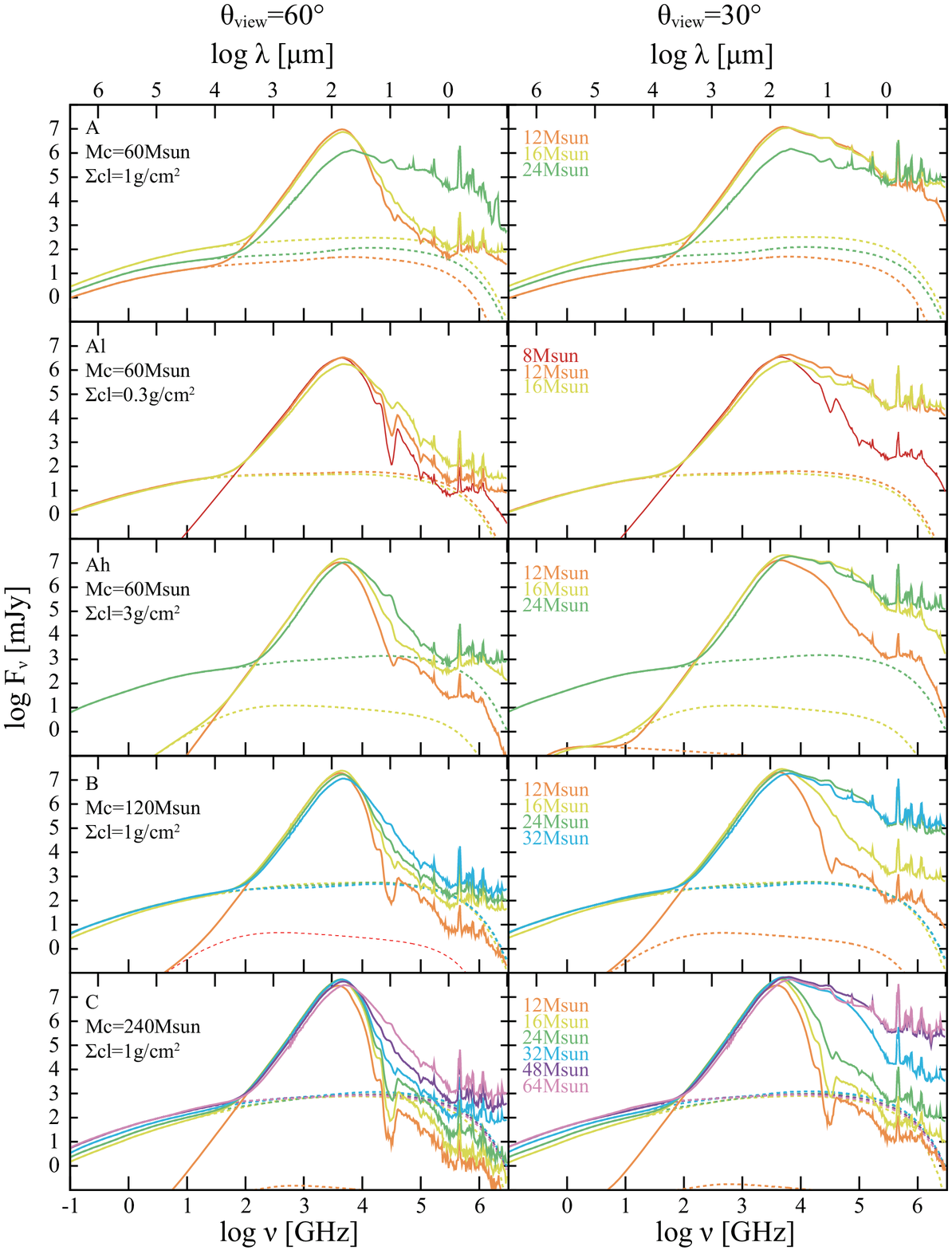}
\end{center}
\caption{
The SEDs at inclination angles of $60\degr$ (left) and $30\degr$
(right) for model group A, Al, Ah, B and C (from top to bottom).  The
dotted lines show the free-free emission components.  }
\label{fig_SED}
\end{figure*}

Figure \ref{fig_SED} shows SEDs for all models.  The features of the
free-free emission are similar to those discussed for model A16:
the free-free fluxes at radio frequencies are almost independent of
viewing angle, and
are $\sim20$ -- $200\:{\rm mJy}$ with
the spectral indices of $p\simeq0.4$ -- $0.7$ at $10\:{\rm GHz}$.
These common
features are caused by the fact that the outflows of all models have
similar structures, i.e., the amount of gas and the density gradient.
If the photoionized region is small, the free-free emission is not
observable, as in, e.g., model Al08.  Once the photoionized region
breaks out, the free-free emission becomes bright, as in, e.g., model
Al12.  Since the flux of free-free emission depends on the density
structure of the outflow, the mass of the outflow that is emitting
free-free flux (within a given distance from the protostar) may be
estimated from its free-free flux.

With our focus in this paper on radio wavelengths, we note again that
we have not accounted for the free-free emission for the total flux at
wavelengths less than $100\:\micron$.  However, in some cases the
free-free flux might make a significant contribution in this range:
e.g., at wavelengths shorter than $10\micron$ of model C16 viewed at
$60\degr$.  The potential importance of free-free emission at shorter
wavelengths will be explored in a future paper.

\subsubsection{Images}\label{sec_image}

Next, we present images of the free-free emission. 
We show both resolved images and those after convolving with finite
beam sizes of various telescopes. Detailed comparisons of models with
individual sources are deferred to future papers in this series.

\begin{figure}
\begin{center}
\includegraphics[width=90mm]{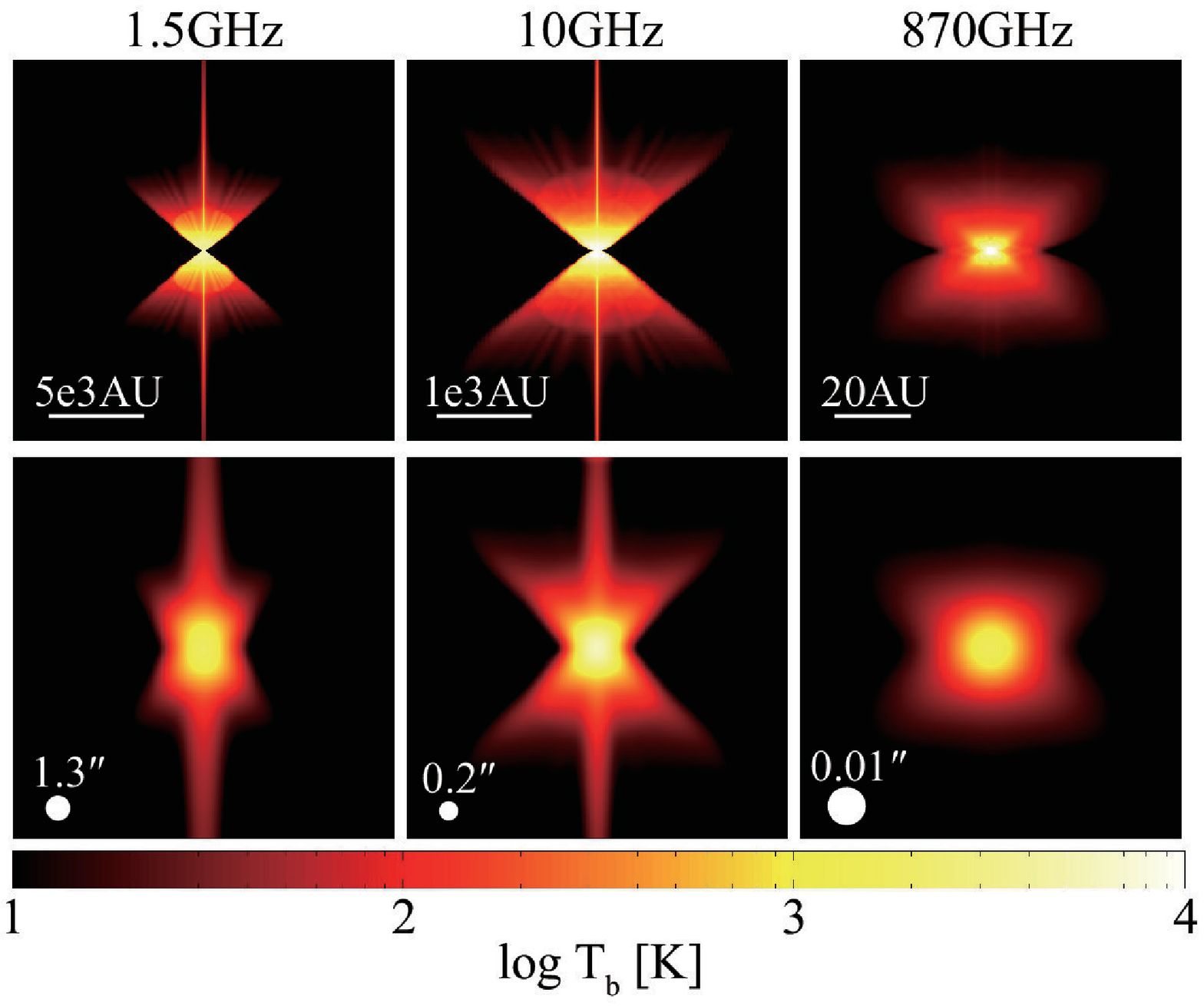}
\end{center}
\caption{
Resolved (top) and convolved (bottom) images of model A16
($M_{c}=60\:\msun, \Sigcl=1\:\gcm,{\rm~and~}m_{*}=16\:\msun$) at an
inclination angle of $60\degr$.  The panels show the images at the
frequencies of $1.5,~10,{\rm~and~}870\:{\rm GHz}$ from left to right.
In the top panels, the box sizes are $20,000{\rm AU}\times20,000{\rm
  AU}$, $4,000{\rm AU}\times4,000{\rm AU}$, and $100{\rm
  AU}\times100{\rm AU}$, respectively.  In the bottom panels, the beam
sizes (FWHM) at each frequency are indicated at the bottom-left
corners ($1.3,~0.2,$ and $0.01\arcsec$).  A distance of $1{\rm\:kpc}$
is assumed.  In the convolved images of the bottom row, from left to
right the brightness temperature of $T_{b}$ is equivalent to
$3.1\mu{\rm Jy} (T_{b}/{\rm K}){\rm\:beam}^{-1}$, $3.3\mu{\rm Jy}
(T_{b}/{\rm K}){\rm\:beam}^{-1}$ and $62\mu{\rm Jy} (T_{b}/{\rm K}){\rm\:beam}^{-1}$ at these given
frequencies.}
\label{fig_img_nu}
\end{figure}

The top three panels of Figure~\ref{fig_img_nu} show the resolved
free-free emission images of model A16 at the inclination of $60\degr$ at three
different frequencies, 
1.5, 10, and 870~GHz. These frequencies correspond to the L and X
bands of the Karl G. Jansky Very Large Array (J-VLA) and band 10 of the Atacama Large
Millimeter/submillimeter Array (ALMA).  Box sizes are $20,000\:{\rm
  AU}\times20,000\:{\rm AU}$, $4,000\:{\rm AU}\times4,000\:{\rm AU}$,
and $100\:{\rm AU}\times100\:{\rm AU}$, respectively.  The apparent
size of the free-free images is larger at lower frequencies since
free-free absorption is stronger (eq. \ref{eq_kapff}).  The
  effective radius at which half of the total flux is emitted is
  $10^{-3} (\nu/10{\rm\:GHz})^{q} \:{\rm pc} \simeq 2\times10^{2}
  (\nu/{10\rm\:GHz})^{q} \:{\rm AU}$, with $q\simeq-0.7$ for
  $1$--$100{\rm\:GHz}$. 
Thus these would typically be categorized as HC \HII~regions (see also
\S\ref{sec_comparison}).  At 1.5~GHz and 10~GHz, the apparent
shape is similar to an hour-glass shape with a jet, because the disk
wind has a self-similar structure at the scale of $\ga1000\:{\rm AU}$.
The surface brightness is high in the central wind region and at the
jet axis where the density is high, and the maximum brightness
temperature of $T_{b}\simeq10^4\:{\rm K}$ indicates optically
thick conditions.  At the higher frequency of 870~GHz, the shape
is now more rounded in the inner region.  This is because of the
accretion disk with radius of $\sim100{\rm AU}$.  The collimated
on-axis jet is not seen, since it becomes optically thin at this
frequency.  For the total free-free flux in Figure~\ref{fig_SEDa16},
the innermost wind region dominates the on-axis jet since its area is
bigger.  However, the bright linear feature is still very interesting,
since it is potentially related to observed radio jets
\citep{gib03,rod08,guz12,guz14}.

The bottom three panels of Figure~\ref{fig_img_nu} show the free-free
images convolved with the finite beam sizes of the telescopes at these
frequencies.  The full-width at half-maximum (FWHM) beam sizes are
1.3,~0.2, and 0.01~$\arcsec$ for 1.5,~10, and 870~GHz, respectively,
i.e., the highest resolutions in the L and X bands of the J-VLA and in
band 10 of ALMA. A distance of 1~kpc is assumed, and the point-spread
function is approximated as a Gaussian function.  The jet feature with
high brightness temperature along the axis is diluted at 1.5~GHz and
8.4~GHz, since it is narrower than the beam size.  However, the linear
feature is still observable at the level of $T_{b}\la30{\rm\:K}$
or $100\mu{\rm Jy\:beam}^{-1}$ at both frequencies. The sensitivity
with root-mean-square value of $\sim10\mu{\rm Jy\:beam}^{-1}$ can be
attained with an observing time of 1~hour and a bandwidth of 500~MHz
by the J-VLA.  The inner brightest region with $T_{b}\simeq10^4{\rm
  \:K}$ is smaller than the beam size, and thus the highest brightness
temperature decreases to $1600{\rm\:K}$ at 870~GHz. The dust emission
cannot have a brightness temperature above the dust sublimation
temperature of $1400\:{\rm K}$.  Therefore, even though the total flux
is dominated by dust emission at the frequencies $\ga100{\:\rm GHz}$
(Fig. \ref{fig_SEDa16}), the free-free emission may be observable even
at this high frequency.  Since most massive protostars are at somewhat
larger distances, this emission may need to be observed in more
massive, denser sources.

\begin{figure*}
\begin{center}
\includegraphics[width=180mm]{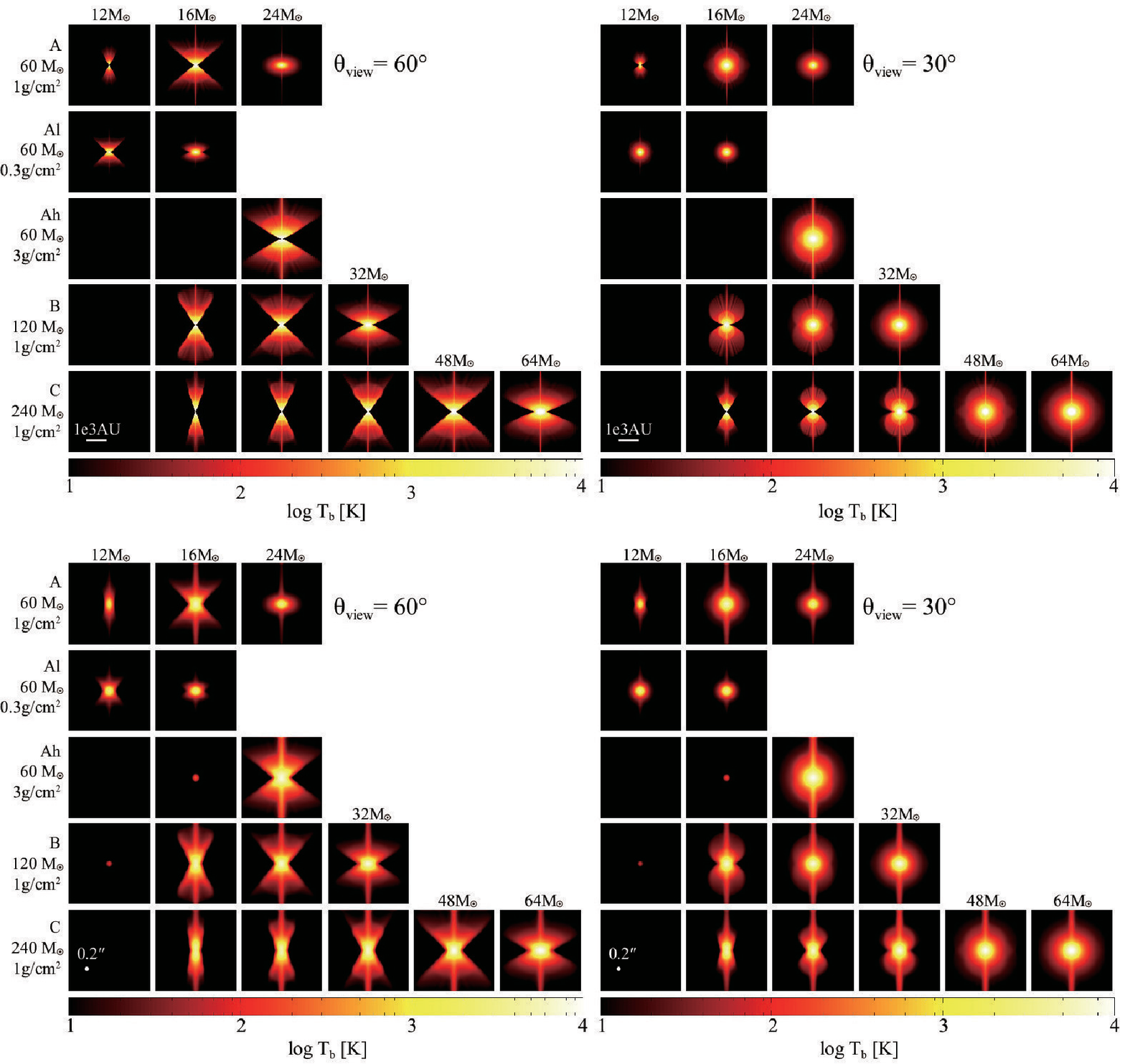}
\end{center}
\caption{
Resolved (top) and convolved (bottom) images at 10~GHz of all models
at inclination angles of line of sight of $60\degr$ (left) and
$30\degr$ (right) to the outflow axis.  The box size of top panels are
$4,000{\rm\:AU}\times4,000{\rm\:AU}$. The images in the bottom panels
are convolved with a beam size of $0.2{\arcsec}$ and a distance of
1~kpc is assumed. Here, the brightness temperature of $T_{b}$ is
equivalent to $3.3\mu{\rm Jy} (T_{b}/{\rm K}){\rm\:beam}^{-1}$.}
\label{fig_img}
\end{figure*}

Figure~\ref{fig_img} shows resolved and convolved images at
  10~GHz as the protostar evolves, viewed at inclination angles of
$30\degr$ and $60\degr$ for all models.  From the resolved images
  (top panels of Fig. \ref{fig_img}), it can be seen that the outflow
opening angle gradually increases, especially at the inclination of
$60\degr$, once the photoionized region is formed.  When the line of
sight passes through the outflow cavity ($\theta_{\rm
  view}<\theta_{\rm w,esc}$), like model A16 at $30\degr$ inclination,
then the image becomes elliptical rather than hourglass shaped.  The
outflow axis always exhibits a high brightness at this frequency.
As stated above, the linear jet-like feature is observable at the
level of $T_{b}\la30{\rm\:K}$ in the convolved image of model
A16.  In the case of higher-$\Sigcl$ or $M_c$, the brightness
temperature is higher since the outflow rate is higher. Thus
the jet feature is observable even at $T_{b}\simeq100{\rm\:K}$
in these models, such as models Ah24 and C32.  The small spots in
convolved images of models Ah16 and B12 are the sign of unresolved
outflow-confined \HII~regions.

\subsection{Hydrogen recombination lines}

\begin{figure}
\begin{center}
\includegraphics[width=70mm]{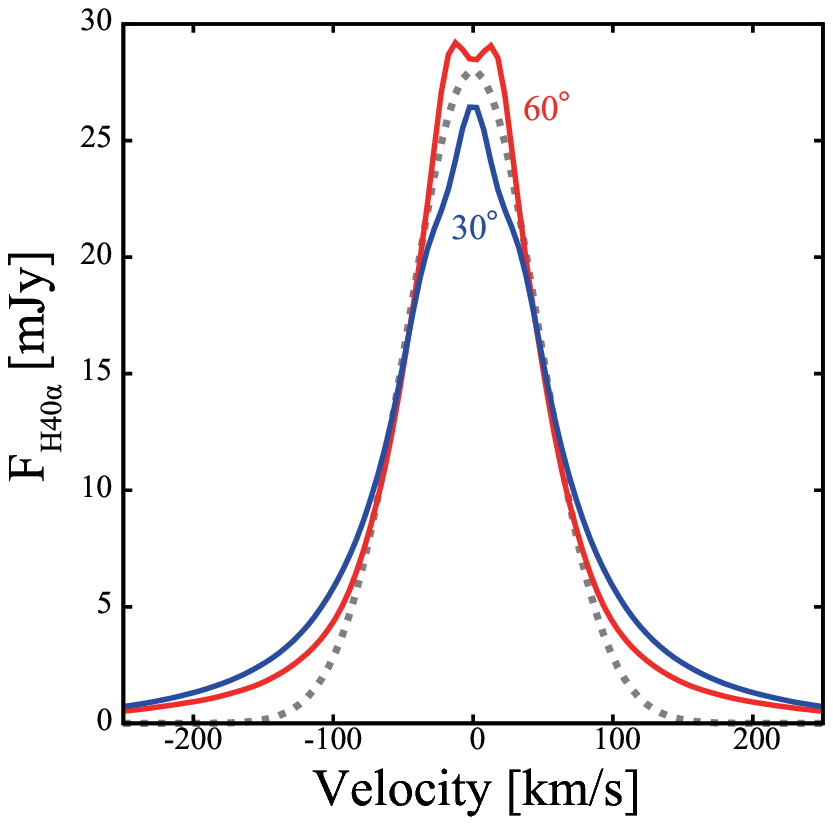}
\end{center}
\caption{
H$40\alpha$ line profiles of model A16 viewed with inclination angles
of $60\degr$ (red solid line) and $30\degr$ (blue solid line).  The
FWHM are $v_{\rm FWHM}=104,\:115\:{\rm km\:{s}^{-1}}$ at
$60,\:30\degr$ respectively.  The gray dotted line shows a Gaussian
distribution with $v_{\rm FWHM}=110\:{\rm km\:s^{-1}}$ and a peak
flux of $28\:{\rm mJy}$.  A distance of $1\:{\rm kpc}$ is assumed.  }
\label{fig_A16lineprofile}
\end{figure}

We perform radiative transfer of the H$40\alpha$ line to obtain
simulated observations that probe the kinematics of the photoionized
wind. Here, a distance of $1\:{\rm kpc}$ and aperture diameter of
$20\arcsec$ (20,000~AU) are assumed.

Figure~\ref{fig_A16lineprofile} shows profiles of H$40\alpha$ flux
from model A16.  The line profiles are very similar for the
inclination angles of $60\degr$ and $30\degr$: the FWHM are
$104{\rm~and~}115\:{\rm km\:s^{-1}}$, the peak fluxes are
$29{\rm~and~}27\:{\rm mJy}$, the ratios of line flux to the free-free
continuum flux are $0.16$ and $0.18$, respectively.  Although we used
a Gaussian line profile of $\phi_\nu$ at each position for the
radiative transfer calculation (eq. \ref{eq_lineprofile}), the
obtained total spectrum has broader wings. This is because the disk
wind has higher velocity components, $\ga500\:{\rm km\:s^{-1}}$, near
the axis.

\begin{figure}
\begin{center}
\includegraphics[width=80mm]{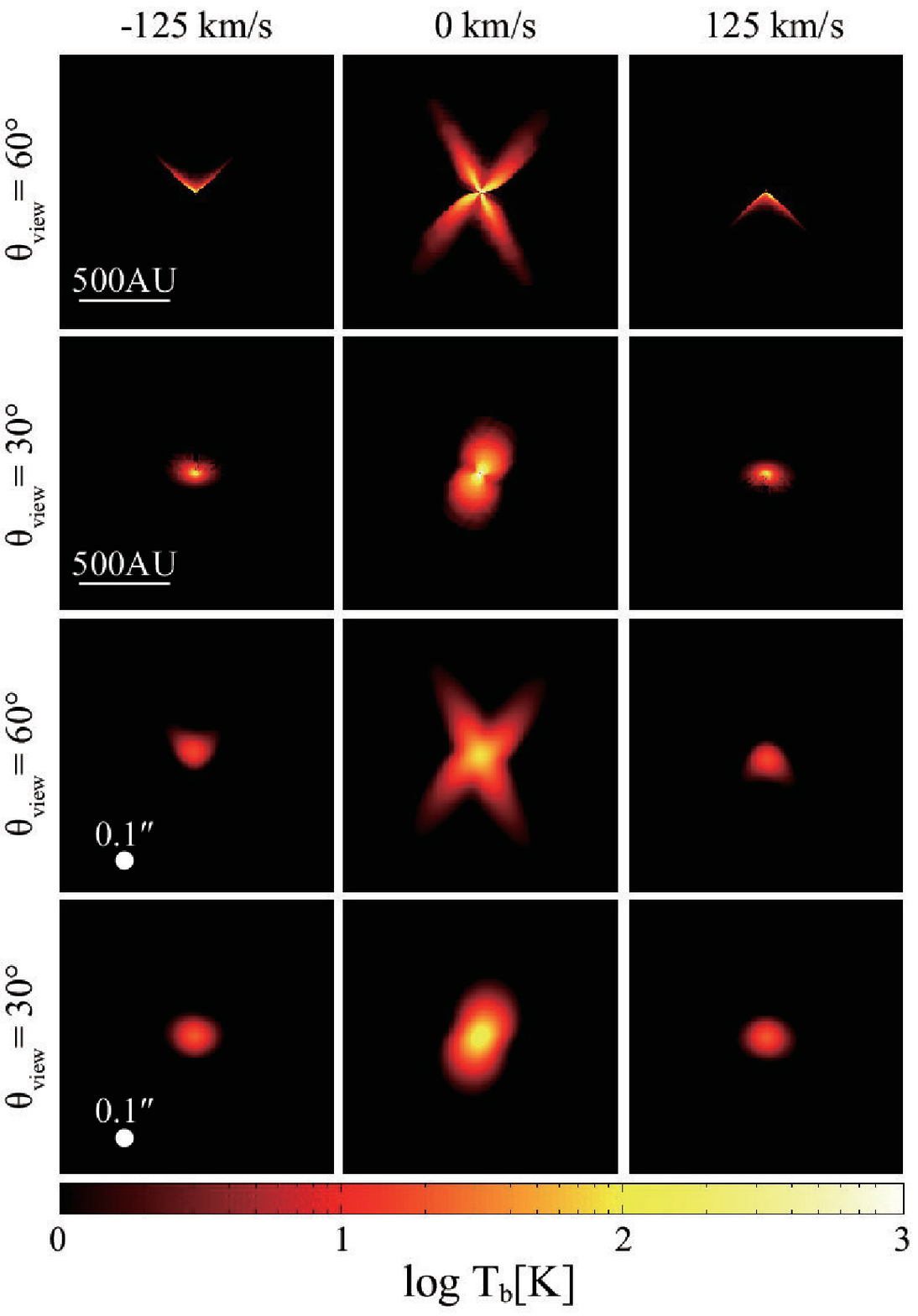}
\end{center}
\caption{
H$40\alpha$ maps of model A16 at $-125,~0,{\rm~and~}125\:{\rm
  km\:s^{-1}}$ (from left to right).  Top two rows show the resolved
images viewed at inclination angles of $60\degr$ (first row) and
$30\degr$ (second row).  The box size is $1500\:{\rm
  AU}\times1500\:{\rm AU}$.  Bottom two rows show the convolved images
viewed at inclination angles of $60\degr$ (third row) and $30\degr$
(fourth row). A beam size of 0.1$\arcsec$ and a distance of 1~kpc are
assumed.}
\label{fig_A16linemap}
\end{figure}

Figure~\ref{fig_A16linemap} shows resolved and convolved maps of
H$40\alpha$ of model A16 at three different velocities,
$-125,~0,{\rm~and~}125\:{\rm km\:s^{-1}}$, at inclinations of
$60\degr$ and $30\degr$.
The map is convolved with a beam size
of 0.1 {\arcsec} which is the highest resolution in band 3 of ALMA.
For this ionized bipolar outflow, the blueshifted component appears on
the north (upper) side and the redshifted component in south (lower)
side.  However, the $0{\rm\:km\:s^{-1}}$ maps are tilted slightly to
northwest-southeast direction.  This is due to the rotation of the
outflow.  Rotating outflows have been found around low-mass and
intermediate-mass protostars \citep{kla13,cod14}, but not yet
for high-mass objects.  In resolved maps at an inclination of
$60\degr$, the line of sight is outside the cavity and thus the rim of
the outflow is brightest.  On the other hand, in resolved maps at an
inclination of $30\degr$, the line of sight passes through the cavity
and the maps are elliptical illustrating the outflow cone.
This difference becomes less clear in the convolved maps
at $\pm125{\rm\:km\:s^{-1}}$.
However, it is still visible in the $0{\rm\:km\:s^{-1}}$ maps at the level of $T_{b}\simeq10{\rm K}$.
The sensitivity of $T_{b}=2{\rm K}$ could be attained with an observing time of
8 hours with a channel width of 6 MHz by ALMA.
The effective spectral resolution is two times of the channel width
which corresponds to 12 MHz or $36{\rm km\:s^{-1}}$ (the H40$\alpha$ HRL is at 99 GHz).
We should mention that again a distance of 1 kpc is assumed here,
and most massive protostars are typically at larger distances.
Thus, this emission may need to be observed in more massive and
denser sources.


\subsection{Escape fraction of ionizing photons}\label{sec_esc}

\begin{figure}
\begin{center}
\includegraphics[width=90mm]{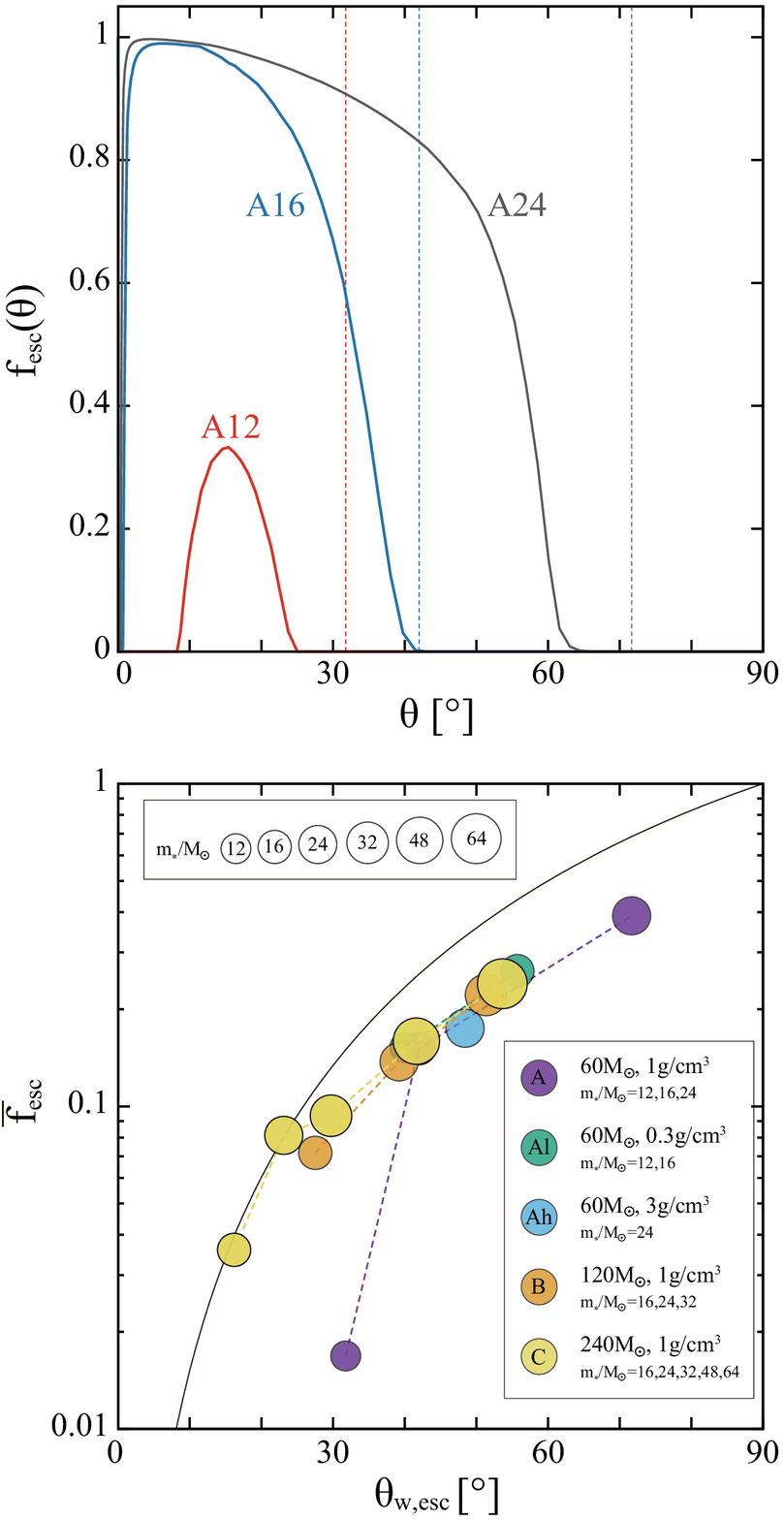}
\end{center}
\caption{
{\it Top}: Escape fraction of ionizing photons to each polar angle
for models A12, A16, and A24 (red, blue, and black solid lines,
respectively).  The dotted lines represent the opening angle of each
model.  {\it Bottom}: The total escape fraction of ionizing photon vs.
the opening angle of the disk wind cavity.  The circles show the
models and the dotted lines represent their evolution (purple: model
group A; green: Al; light blue: Ah; orange: B; and yellow: C).
The size of each circle represents the protostellar mass. 
The solid line shows the covering fraction of the whole sky solid angle by
the disk wind cavity, i.e., $1-\cos\theta_{\rm w,esc}$.
}
\label{fig_escape}
\end{figure}

So far we presented results for the ionization of disk winds.  Next,
we consider the ionizing photons that escape along the outflow
cavities created by these winds.  Since the obtained ionized regions
are unconfined in the radial direction, some fraction of ionizing
photons escape and are expected to ionize the ambient clump gas.  This
ionization feedback, in addition to that of the disk wind protostellar
outflows, may affect star formation around the forming massive star
and will also lead to additional radio emission.

The top panel of Figure~\ref{fig_escape} shows the escape fraction of
ionizing photons to each polar angle, $f_{\rm esc}(\theta)$, along
with the opening angle of the disk wind cavity, $\theta_{\rm w,esc}$.
In the case of model A12, the escape fraction is $0.35$ at its maximum
value and almost zero for $\theta<10\degr$ or $\theta>30\degr$.  This
is because the ionized region is restricted to
$10\degr\la\theta\la30\degr$ (Fig.~\ref{fig_T_a}).  As the ionized
region expands in models A16 and A24, the angular range in which
photons escape also expands and the maximum value of $f_{\rm
  esc}(\theta)$ reaches about unity.  However, the escape fraction is
still $<0.5$ in directions near the boundary of the core infall
envelope (i.e., $\theta>\theta_{\rm w,esc}-10\degr$: see also
Fig. \ref{fig_diff}) or near the outflow axis ($\theta<5\degr$: see
also Fig. \ref{fig_log_A16}) due to strong attenuation in these higher
density regions.

The bottom panel of Figure~\ref{fig_escape} shows the evolution of the
outflow cavity opening angle, $\theta_{\rm w,esc}$, and the total
escape fraction of ionizing photons, $\overline{f}_{\rm esc}$, for all
model groups.  It can be seen that the total escape fraction increases
as the opening angle grows in every case.  However, we note that, due
to Hydrogen ionization and dust absorption, the total escape fraction
is lower than the fraction of solid angle extended by the disk wind
cavity.  The maximum value of $\overline{f}_{\rm esc}$ in our models
is $0.38$ with $\theta_{\rm w,esc}=72\degr$, so more than half of
ionizing photons do not escape from the system (including the
directions blocked by the disk and infalling core envelope).

The total ionizing photon rate escaping from the protostellar core is
$\overline{f}_{\rm esc} S_{\rm *,acc}$.  Most of these photons will be
able to ionize the ambient clump or larger-scale cloud gas, especially
the gas in front of the head of the outflow bowshock.  Our model does
not explicitly include the structure of the outflow bowshock or the
ambient clump gas.  Thus, we do not study details of the geometry or
kinematics of the ionized clump.  Instead, as a simple first
approximation, we use a Str\"omgren sphere analysis to make some
simple estimations. In this case, the approximate size scale of the
ionized region created by escaping photons is
\begin{eqnarray}
	R_{\rm St,cl} = \left( \frac{3 \overline{f}_{\rm esc} S_{*,\rm
            acc}}{4\pi n_{\rm H,cl}^2 \alpha_{\rm B}}
        \right)^{1/3}, \label{eq_r_cl}
\end{eqnarray}
where $n_{\rm H,cl}$ is the density of the ambient clump, and
$\alpha_{\rm B}=\alpha_{\rm A}-\alpha_{1}$ is the case B recombination
rate (this estimate ignores absorption of photons by dust).  The mean
clump density is given by $n_{\rm H,cl} = 2.0\times10^5
({\Sigcl}/{1\:{\gcm}})^{3/2} (M_{\rm cl}/4000\:M_\odot)^{-1/2} {\:\rm
  cm^{-3}}$.
The recombination rate, $\alpha_{\rm B}$, is a function of
temperature, which is calculated from the stellar spectrum and the
clump density (\S\ref{sec_T}).  The free-free emission is evaluated by
solving equation (\ref{eq_RTff}) assuming spherical geometry.
The FWHM of the velocity distribution of the ionized clump gas,
which will be used as part of the diagnostic process to classify \HII~regions
(\S\ref{sec_discussion}), is estimated simply from thermal broadening,
$v_{\rm FWHM,cl} = \sqrt{8\ln{2}} \sigma_{v}$.  This estimation gives
the minimum value of the line width, since the clump gas could also be
disturbed by outflows and additional turbulent motions that broaden
the line profile.

We note some caveats of this Str\"omgren sphere approximation for
ambient clump ionization.
First, the geometry of the ionized region is
likely to be affected by the outflow cavity, leading to deviations from
a simple sphere.
Second, the ambient clump density is expected to be inhomogeneous,
which can lead to various effects, including enhanced effective
densities at ionization fronts that can lead to smaller \HII~region
sizes \citep[e.g.,][]{tan04b}.
Thus, the free-free flux evaluated by the Str\"omgren sphere analysis
is only expected to be a very crude approximation.  If the ambient
clump region is optically thin, its total free-free flux from the
ionized region is just proportional to the total escaping photon rate,
$\overline{f}_{\rm esc} S_{\rm *,acc}$, regardless of its shape or
density. However, if the ionized region is optically thick, the
free-free flux is smaller than the optically thin limit by the factor
of its mean optical depth. Therefore, the error in the free-free flux
would be small if the ionized clump region is optically thin.


\section{Discussion} \label{sec_discussion}

\subsection{Outflow-confined \HII~regions and their break out}\label{sec_confined}

Here we summarize the evolution of ionized outflows. 
Note, the increase of ionizing flux in the Kelvin-Helmholz contraction
phase is very rapid (Fig. \ref{fig_protost}) and our numerical results
have only been carried out with relatively coarse sampling across the
evolutionary sequence. However, we can still present a description of
the following evolutionary sequence.
At the first stage, when the ionizing flux is small, the MHD-driven
outflow is the first barrier to the propagation of ionizing photons,
except in the very narrow range of directions within the inner disk
wind cavity. The \HII~region is confined in a very thin layer of the
inner outflow wall, e.g., model Ah16 in Figure \ref{fig_ah}.
As the ionizing flux increases, the \HII~region extends to greater
distance from the protostar.
At a critical ionizing flux, the \HII~region breaks out at a middle latitude
(see model A12 in Fig \ref{fig_T_a}).
At the lower latitude, the density is higher than that in the middle latitude
since it is closer to the launching zone of the disk wind.
On the other hand,
the density gradient at the higher latitude is flatter than that at the middle latitude due to the collimated structure of the wind.
Therefore,
the \HII~region breakout starts from $\theta \simeq\theta_{\rm w,esc}/2$.
The critical ionizing flux for the
radial breakout depends on the detailed structure of outflow, i.e.,
mass loss rate, opening angle and velocity structure.
From a qualitative standpoint, the mass loss rate is roughly proportional to
the mass accretion rate, and thus the critical ionizing flux for the
breakout is higher for higher-$\Sigcl$ cases.  The breakout does not
occur at $8\:\msun$ even in the low-$\Sigcl$ case of model Al08.  The
typical flux for breakout is $10^{46}$--$10^{47}\:{\rm s}^{-1}$ in our
numerical calculations, which is as high as the flux from the ZAMS at
about $10$--$15\:\msun$.  All models in this work reach $20\:\msun$ at
the final stage, however, it should be noted that there are almost as
many stars between $8$ and $10\:\msun$ as above $20\:\msun$.  In the
formation of such stars with $\lesssim 10\:\msun$, the \HII~region is
confined during the main accretion phase and breakout occurs only
towards the end of accretion when the outflow rate also decreases.

After first breakout at $\theta_{\rm w,esc}/2$, the photoionized
region extends in polar angle. To ionize almost the entire outflow, the
ionizing flux needs to increase by about one order of magnitude from
the first breakout ionizing flux. This takes only $10^3$--$10^4{\:\rm
  yr}$ since the ionizing flux increases rapidly in the KH contraction
phase.  Once the entire outflow is ionized, the expansion in polar angle of
the \HII~region slows down matching the increase of the outflow cavity
opening angle, which has a timescale similar to the accretion time,
i.e., $m_*/\mdot_*\sim10^{4}$--$10^{5}{\:\rm yr}$.

Our model of photoionized disk winds may also help to explain the
short timescale variability that is seen in some UC/HC \HII~regions
\citep{fra04,rod07,gal08} and which has also been modeled in the
radiative hydrodynamical simulations of \citet{pet10}. Fluctuations in
the accretion rate would also lead to changes in the outflow massloss
rate, which would change the density and thus the extent of the
photoionized region. Changes could happen on outflow crossing
timescales, i.e., as short as $\sim 0.01$~pc / $1000\:{\rm km\:s}^{-1}
\simeq 10$~yr.

Our model makes prediction for the spectral and the size indices,
i.e., $p$ and $q$, of the free-free emission.
The free-free emission spectral index at 10~GHz is about
 $p=0.4$--$0.7$  (see \S 3.2.1),
which indicates that the outflow is partially optically thick.
The apparent size of the observed free-free emission (see \S 3.2.2)
is as small as
$\sim10^{-3}{\rm\:pc}$ at 10 GHz
and smaller at higher frequencies,
scaling with size as $r\propto \nu^q$, 
with the index of $q\simeq-0.7$, even though the ionized region is
larger than the core scale of $0.1\:{\rm\:pc}$ after breakout.  These
frequency dependences are related to the density structure of the
ionized outflow.  A simplest analytical structure is a spherical wind
with constant velocity, i.e., $n_{\rm H} = \mdot_{w}/(4\pi r^2
v_{w})$ where $\mdot_{w}$ and $v_{w}$ are the wind mass
loss rate and velocity \citep{pan75}.  Since the opacity for the
free-free emission at radio frequencies is approximatly $\kappa_{\nu,
  \rm ff}\propto n_{\rm H}^2 \nu^{-2.1}$ if the gas is isothermal, the
optical depth rapidly decreases as $r^{-3}$.  Therefore, the outer
region becomes optically thin and has less contribution to the total
free-free flux even though the ionized gas extends to larger scales.
In this spherical model, the radius of the optically thick region
scales as $r_{\tau_\nu=1}\propto \nu^{-0.7}$, and thus the total
free-free flux, which is proportional to the flux from the optically
thick region, is $F_{\nu, \rm ff}\propto B_{\nu}r_{\tau_\nu=1}^2
\propto \nu^{0.6}$.
Additionally, as shown by \citet{rey86}, even with a bipolar jet
geometry, the set of indices will be similar to the simple spherical
wind if the jet is conical (constant aperture angle), of constant
velocity and ionization fraction, and is isothermal. Thus the
departure from the indices of $(p,\:q)=(0.6,\:-0.7)$ is related to
acceleration, collimation, thermal structure, or variation of
ionization fraction \citep{rey86, guz14}. Our wind model is more
complex compared to the spherical wind or the conical jet,
however the fundamental physics is the same and thus the values of its
spectral and size indices are similar.

\subsection{Comparison with observed radio sources around massive protostars}\label{sec_comparison}

Here, we compare general properties of our model with observed radio
sources around massive protostars and show that the features of our
theoretical models are similar to those of observed radio winds and
jets.

\begin{figure*}
\begin{center}
\includegraphics[width=180mm]{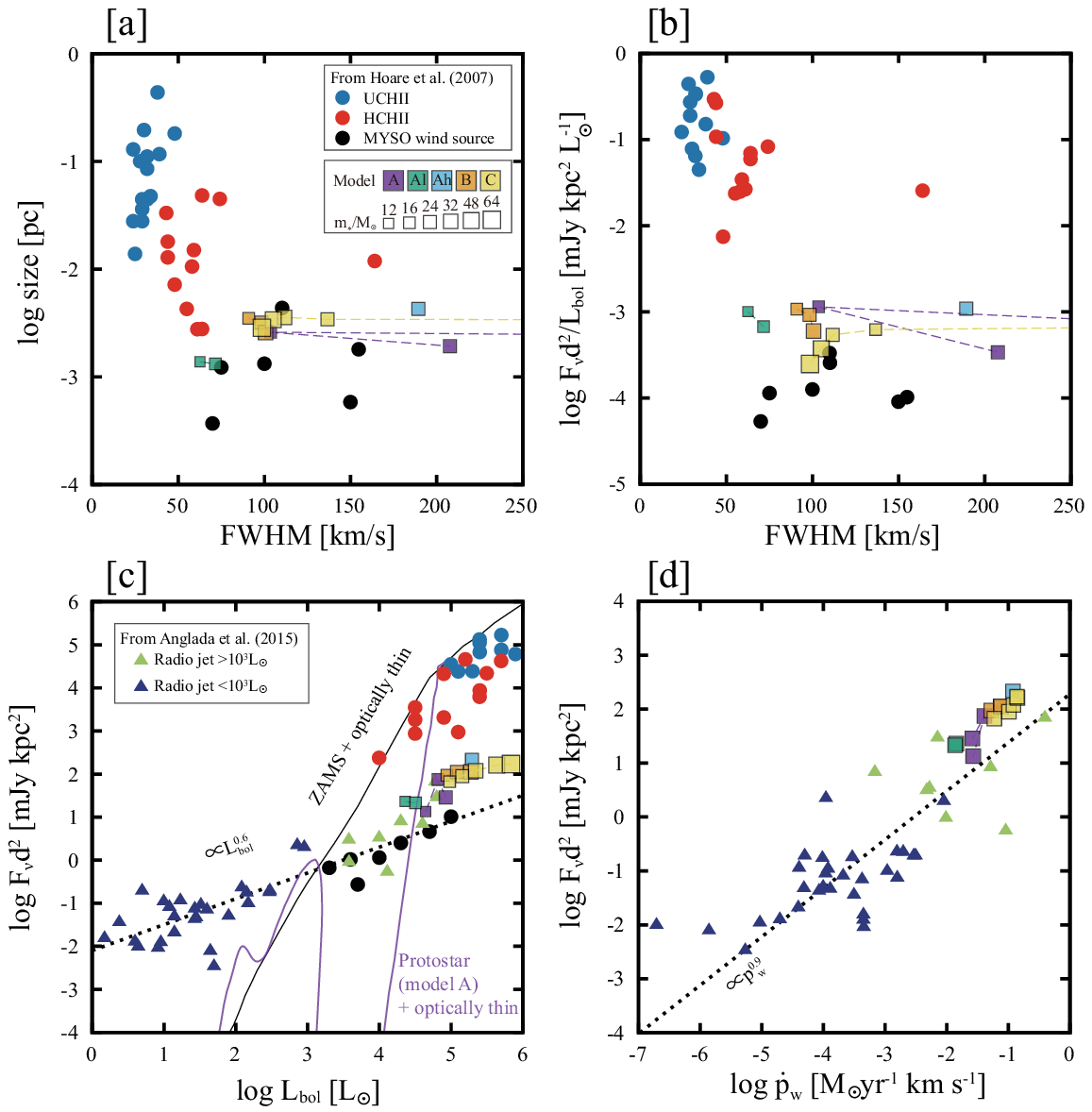}
\end{center}
\caption{
Properties of theoretical outflow-confined \HII~region models compared
with observed radio sources around protostars.  The observational data
are adopted from \citet{hoa07}, \citet{hoa07b} (filled circles in
panels (a), (b), and (c)), and from \citet{ang15} (filled triangles in panels
(b) and (c)).  In all panels, the theoretical models are shown by square
symbols with sizes indicating the stellar mass $m_*$ (purple: model
group A, green: Al, light blue: Ah, orange: B, and yellow: C).  The
dash lines represent evolution within the model groups.  (a): Size
versus line width.  Observed objects are shown by circles (blue: UC
\HII~regions, red: HC \HII~regions, and black: MYSO wind sources).
The sizes for the theoretical models (which vary with frequency) and
the observed sources have both been evaluated at $8\:{\rm GHz}$.  (b):
The ratio of radio and bolometric luminosities versus line width.
Symbols are same as the panel (a).  The radio luminosity for the
theoretical models, which varies with frequency, and the observed
sources have been evaluated at $8\:{\rm GHz}$.  (c): Radio luminosity
versus bolometric luminosity.  Triangles show observed radio jets
(light green: high-luminosity, and blue: low-luminosity sources).  The
radio luminosities of the theoretical models are evaluated at $8\:{\rm
  GHz}$, and the observational data are at $5$ or $8\:{\rm GHz}$.  The
dotted line shows the fitting by \citet{ang15}: $F_\nu d^2/{\rm
  mJy\:kpc^2} =8\times10^{-3}\left( L_{\rm bol}/L_\odot
\right)^{0.6}$.  This relation is based on observed radio jets over a
wide bolometric luminosity range of $1$--$10^5L_\odot$.  The solid
line shows the estimated radio flux from an optically thin \HII~region
based on the ionizing luminosities of ZAMS stellar models by
\citet{tho84}.  The purple line shows the estimated radio flux from an
optically thin \HII~region based on the ionizing luminosities of the
protostar of model group A.  (d): Radio luminosity versus outflow
momentum rate.  The dotted line shows the fitting by \citet{ang15}:
$F_\nu d^2/{\rm mJy\:kpc^2}
=190\left({\dot{p}_w}/{\msunyr{\rm\:km\:s^{-1}}} \right)^{0.9}$.  The
square and triangle symbols and the purple line are same as in panel
(c).  The momentum rate of our model $\dot{p}_w$ is obtained from the
evolution calculation by ZTH14 rather than from estimation by a
synthetic observation.  Properties of the theoretical models of
outflow-confined \HII~regions are broadly consistent with those of the
high luminosity end of the observed radio wind/jet sources.  }
\label{fig_comparison}
\end{figure*}

Figure \ref{fig_comparison} shows the properties of size, line width,
radio and bolometric luminosities, and outflow momentum rate of
our models and of observed radio sources.  For our models, size and
free-free luminosity are measured at $8\:{\rm GHz}$ and line width is
evaluated from H$40\alpha$ assuming an inclination angle of $60\degr$.
The size is evaluated by the diameter at which half of the total radio
flux of the system is emitted.  As we have seen, the inclination angle
does not affect these properties significantly at radio frequencies.
Note, the outflow momentum rate, $\dot{p}_w$, of our model shown
in Figure \ref{fig_comparison} is obtained from the protostellar
evolution calculation of ZTH14 rather than from an estimation via a
synthetic observation.

In panel (a) of Figure \ref{fig_comparison}, we show how \citet{hoa07}
distinguished objects into three categories based on size and line
width: UC \HII~regions are $\sim0.1{\rm\:pc}$ and $<40\:\kms$, HC
\HII~regions are $<0.05{\rm\:pc}$ and $>40\:\kms$, and massive young
stellar object (MYSO) wind sources are $<0.005{\rm\:pc}$ and
$>70\kms$.  It should be noted that this classification of the radio
sources is somewhat arbitrary.  For example, \citet{kur02,kur05}
defined UC \HII~regions as having diameters $\leq0.1{\rm\:pc}$ and HC
\HII~regions as $\leq0.01{\rm\:pc}$ with emission measures
$\geq10^8{\rm\:pc\:cm^{-6}}$.  \citet{tan14} adopts a simpler
definition only based on the above sizes.  When we refer to data from
\citet{hoa07} and \citet{hoa07b}, we follow Hoare et al.'s
classification.  The line width of UC and HC \HII~regions are derived
from radio HRLs, while those of wind sources are from IR HRLs.  The
wind sources (BN, W33A, S140 IRS 1, NGC 2024 IRS 2, GL 989 and GL 490
are plotted here) are the smallest and fastest objects.  Comparing in
those plots, our models reproduce the properties of wind sources
quantitatively, with sizes of $\sim10^{-3}\:{\rm pc}$ and line widths
of $\sim100\kms$.  Two of our models have broader line widths than
$250\kms$ ($718\kms$ for A12 and $429\kms$ for C16).  These two models
are at the beginning of \HII~region expansion, and the ionized region
is confined only near the axis where the wind velocity is
highest. Thus, their line width is exceptionally broad.

The difference between UC/HC \HII~regions and MYSO wind sources is
more clearly seen in the ratio of radio and bolometric luminosities in
panel (b) of Figure \ref{fig_comparison}.  UC/HC \HII~regions are more
radio-loud than MYSO wind sources by two orders of magnitude or more.
Even though the average radio to bolometric luminosity ratio of
our models are about one order of magnitude higher than those of
MYSO wind sources, there is still overlap given their dispersions.

\citet{ang15} found two empirical correlations related to the radio
luminosity of observed radio jets: one is a correlation with
bolometric luminosity; the other is with outflow momentum rate:
\begin{eqnarray}
	\frac{ F_\nu d^2}{{\rm\:mJy\:kpc^2}} &=&8\times10^{-3} \left( \frac{ L_{\rm bol} }{ L_\odot } \right)^{0.6};\label{eq_LL}\\
	\frac{ F_\nu d^2}{{\rm \:mJy\:kpc^2}}&=&190
	\left( \frac{ \dot{p}_w }{\msunyr{\rm\:km\:s^{-1}}} \right)^{0.9}.\label{eq_pL}
\end{eqnarray}
Interestingly, those two correlations are consistent over a wide
bolometric luminosity range of $\sim 1\:L_\odot$ to
$\sim10^5\:L_\odot$ (dotted lines in panels (c) and (d) of
Fig. \ref{fig_comparison}). MYSO wind sources from \citet{hoa07b}
(black circles) also fall on the first correlation.  The solid
line in panel (b) is the expected radio luminosity in the
optically thin limit using the ionizing photon rate of ZAMS models by
\citet{tho84}, which represents the maximum radio luminosity from
photoionized gas.  At $L_{\rm bol}\la10^{3}\:L_\odot$, however, there
are radio jets that exceed this maximum value \citep{ang92}.  It has
been suggested that the single power law correlations of
equations (\ref{eq_LL}) and (\ref{eq_pL}) indicates a common nature
of (ionized) outflows from low-mass and high-mass protostars
\citep{rod89, ang95}.  Since the photoionization properties differ
greatly from low-mass to high-mass stars, this might imply that a
different mechanism, i.e., shock ionization, is responsible for the
observed radio emission along these correlations.  However, we
find that the radio luminosities of our model agree reasonably well
with both the correlations of eqs. (\ref{eq_LL}) and
(\ref{eq_pL}), even though they are photoionized structures.  We
predict that there should be a transition to photoionization-dominated
winds for massive protostars, but that in this regime the
radio/bolometric luminosities and the outflow momentum rate may not
differ too much from this empirical relation.  
The systematic differences of our models
are towards somewhat higher values of radio luminosities by factors of
1.5--13. This may be a signature of the initial stage of
photoionization leading to enhanced radio fluxes. Note, our models
tend to be more massive than the observed sources, so more data in
this range is needed.

If the photoionized region is optically thin, the free-free emission
is simply proportional to the ionizing photon flux rate \citep{hum63}.
However, our model indicates much lower free-free flux than the
optically thin limit (panel (c) in Fig. \ref{fig_comparison}).
First, this is because more than half of the emitted ionizing photons
are absorbed at the innermost region of $r\ll1000{\rm AU}$.  This
region is optically thick for the radio free-free emission, and thus
the absorption there does not contribute to the total free-free flux.
In the case of model A16, $78\%$ of photons are absorbed inside
$r\leq100{\rm AU}$.  Second, most of the rest of the photons, ${\bar
  f}_{\rm esc}$, escape from the system.  Those escaping photons do
not induce free-free flux from the core scale, but contribute to the
flux from the ambient clump gas, which is discussed below
(\S\ref{sec_ion_clump}).  The final reason is that the
photoevaporation flow has so far not been included in our model.  If
the ionizing photon rate is high enough to ionize the disk and
envelope surfaces, e.g., in the case of model A24, the outflow rate of
ionized gas increases and the free-free flux should be enhanced.
While the first and second reasons are physical, the last one is due
to the incompleteness of our model.  We will address photoevaporation
and its feedback effects in a subsequent paper.

Our ionized outflow models typically have spectral indices with
$p=0.4$--$0.7$ (Fig. \ref{fig_SED}), and size index with frequency of
$q\simeq -0.7$ from free-free emission at 10~GHz
(Fig. \ref{fig_img_nu}).  As discussed above, those indices are
related to the density gradient of the outflow.  The indices obtained
by our calculations are consistent with some observed radio jets, such
as MWC 349 \citep[$p=0.67\pm0.03$ and $q=-0.74\pm0.03$,][]{taf04} and
W75N(B) \citep[$p=0.61\pm0.03$ and $q=-0.7\pm0.03$,][]{car15}.
G345.4938+01.4677, however, has the slightly steeper indices of
$p=0.92\pm0.01$ and $q=-1.1^{+0.5}_{-0.4}$ than our model, which may
be interpreted as being due to a stronger degree of collimation and
acceleration \citep[][Guzm{\'a}n et al., in prep.]{guz14}.

We should notice that, even though our theoretical model has similar
properties to those of the observed radio jets, there is still a lack
of some physical processes in our model.  The velocity of some jets is
estimated to be as high as $500\kms$ from the proper motion of
associated lobes along the axes, e.g., HH 80-81
\citep{mar98,mar95,mas15} and Cep A HW 2 \citep{cur06}.  Similar to
these sources, our wind models have high radio brightnesses and high
velocities of $\ga300\:\kms$ near the axes
(Figs.~\ref{fig_n}, \ref{fig_al}, etc.). However, unlike our
model, some of lobes are found to have the radio emission with
negative spectral indices,
such as Serpens \citep{rod89a}, HH 80-81 \citep{mar93}, Ceph A HW 2
\citep{gar96}, W3(OH) \citep{wil99}, and IRAS 16547-4247
\citep{rod05}. These negative spectral indices can be interpreted as
due to nonthermal synchrotron emission induced by strong shocks with
the ambient gas \citep{aro07}, and this has been confirmed by the
detection of linearly polarized emission in HH 80-81 \citep{car10}.
Additionally, our simple axisymmetric model has not accounted for the
possibility of outflow axis precession, which is apparent in
IRAS16547-4247 \citep{rod08}.  Although the current number of sources
with these features is relatively small, inclusion of such additional
physics in our theoretical models, i.e., shock ionization and outflow
axis precession, are likely to be needed for a complete understanding
of jet properties in massive star formation.  Such analysis is also
deferred to a future paper.

As we have seen, the properties of our models are consistent with the
so-called ``MYSO wind sources" in \citet{hoa07} and \citet{hoa07b},
and ``radio jets" with high bolometric luminosity in \citet{ang15}.
In general, the main difference of ``winds" and ``jets" is the shape,
i.e., winds have relatively wide opening angles and jets are more
collimated bipolar outflows.  This difference could be related to
evolutionary stage (see also Fig. \ref{fig_img}).
To clarify our view of the evolutionary sequence of massive star
formation and its outflow launching mechanism, further study is needed
utilizing multi-wavelength observations to be compared with
theoretical models.

\subsection{Ionization of ambient clump gas by escaping photons} \label{sec_ion_clump}

\begin{figure*}
\begin{center}
\includegraphics[width=180mm]{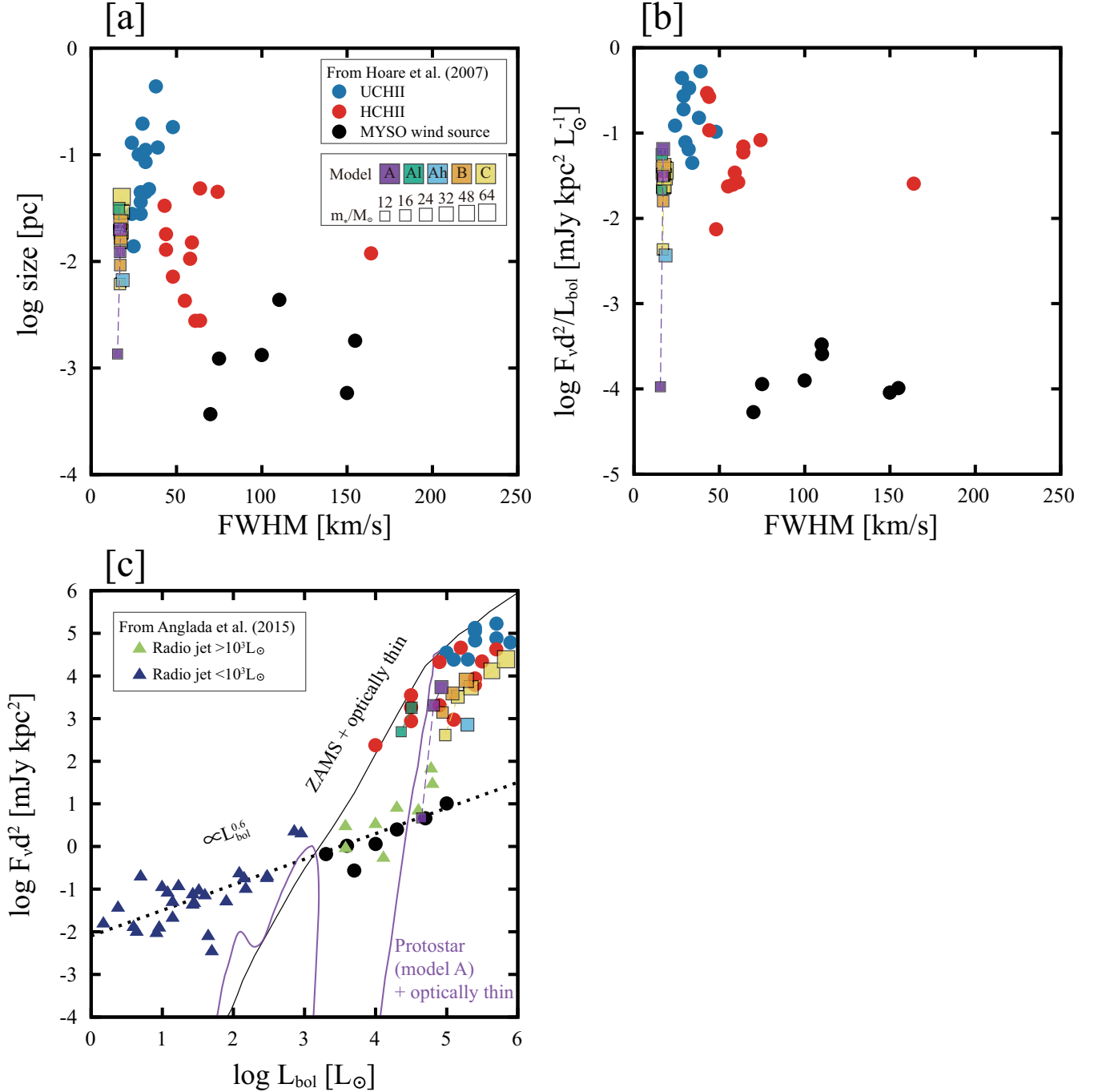}
\end{center}
\caption{
Same as panels (a), (b) and (c) of Figure \ref{fig_comparison}, except theoretical models now
account for the escape of ionizing photons from the disk wind cavity.
}
\label{fig_comparison_b}
\end{figure*}

Finally, we comment on the ionization of the ambient clump.  Tens of
percent of the total ionizing photon output escape along the outflow
cavity (Fig. \ref{fig_escape}).  These photons are likely to ionize
the ambient clump which would lead to enhanced radio luminosity.  We
estimate features of these extended ionized regions, as described in
\S\ref{sec_esc}.
Figure~\ref{fig_comparison_b} shows the comparison of our ionized
clump models and observed sources (the observational data are the same
as in Fig.~\ref{fig_comparison}).  The Str\"omgren-based models of
ionized ambient clumps typically have the size of $\sim10^{-2}\:{\rm
  pc}$ and line widths of $\sim20\kms$, and large radio luminosities
of $\ga10^2\:{\rm mJy\: kpc^{2}}$, which are larger in size, narrower
in line width, and more radio loud than the ionized outflow models.
These features are roughly consistent with observed UC/HC
\HII~regions, except the line widths are broader in the observed
systems. However, this difference is not surprising as the theoretical
models include only thermal broadening and so give minimum widths.

We conclude that some UC/HC \HII~regions may be formed by the escaping
photons from massive protostars with radio jets or winds.  In this
study, the ionized clump features are evaluated based on the simple
Str\"omgren analysis (\S\ref{sec_esc}). For more detailed analysis,
planned for future work, one needs to model more accurately the
structure and kinematics of ambient clump gas, including its
perturbation by the outflow, to better understand the connection of
radio winds/jets and UC/HC \HII~regions.

\section{Summary}\label{sec_conclusions}
We have studied the structure, evolution and observational properties
of photoionized disk wind outflows that are expected to be associated
with massive protostars forming by Core Accretion. Specifically, we
have followed the quantitative properties of the Turbulent Core
Accretion Model of \citet{mck03} and follow-up protostellar evolution,
disk wind and IR continuum radiative transfer models of \citet{zha11}
and \citet{zha13,zha14}.

We calculated the transfer of ionizing photons and obtained the
photoionized structure.  The disk wind starts to be ionized when the
stellar mass reaches about $10$ -- $20\:\msun$, depending on the
properties of the initial cloud core, due to the stellar temperature
increase resulting from Kelvin-Helmholz contraction.  At this moment,
the photoionized region is confined to zenith angles of about $10$ --
$30\degr$, while the high density regions near the outflow axis and
equatorial plane remain neutral. On the other hand, in the radial
direction, the photoionized region is unconfined, since the density
gradient of the disk wind is steep.  The disk wind is almost entirely
ionized in about $10^3$ -- $10^4{\rm yr}$, and the photoionized region
extends as the outflow opening angle increases.  The outer disk
surface is not ionized even though the diffuse radiation has been
considered.  This is because of the shielding effects of the inner
disk and the disk wind.

Using the obtained photoionized structures, we calculated radiative
transfer of free-free continuum and Hydrogen recombination line
emission.  The free-free emission exceeds dust emission at the radio
frequencies lower than $100\:{\rm GHz}$ once the ionized region is
formed.  While shorter wavelength infrared flux from dust emission can
depend on inclination of viewing angle, the radio flux from free-free
emission is almost independent of inclination and
its typical flux density is
$\sim(20\:$--$\:200)\times(\nu/10{\rm GHz})^p{\rm\:mJy}$
at a distance of 1 kpc
with a spectral index $p\simeq0.4\:$--$\:0.7$
The apparent size depends on
frequency as approximately $500(\nu/10{\rm GHz})^{-0.7}\:{\rm AU}$.
The profile of ${\rm H40\alpha}$ has a FWHM of $\sim100\:\kms$ with
broad wings.  These properties of our model are similar to those of
observed radio winds and jets from massive protostars.  Detailed
comparison with individual sources is deferred to future papers in
this series.

\section*{Acknowledgments}
The authors thank Christopher McKee,
Andres Guzm{\'a}n, Taishi Nakamoto, Hiroyuki Kurokawa,
Kohei Inayoshi, and Shuo Kong for fruitful discussions and comments.
The authors also thank the anonymous referee for comments, which were useful to improve the original manuscript.
JCT acknowledges support from NSF grant AST 1411527.


\clearpage


\begin{thebibliography}{}

\bibitem[Anglada et al.(1992)]{ang92}
	Anglada, G., Rodr{\'{\i}}guez, L.~F., Canto, J., Estalella, R., \& Torrelles, J.~M.\ 1992, \apj, 395, 494 

\bibitem[Anglada(1995)]{ang95}
	Anglada, G.\ 1995, Revista Mexicana de Astronomia y Astrofisica Conference Series, 1, 67

\bibitem[Anglada et al.(2015)]{ang15}
	Anglada, G., Rodr{\'{\i}}guez, L.~F., \& Carrasco-Gonzalez, C.\ 2015, Advancing Astrophysics with the Square Kilometre Array (AASKA14), 121 
	
\bibitem[Araudo et al.(2007)]{aro07}
	Araudo, A.~T., Romero, G.~E., Bosch-Ramon, V., \& Paredes, J.~M.\ 2007, \aap, 476, 1289 

\bibitem[B{\'a}ez-Rubio et al.(2013)]{bae13}
	B{\'a}ez-Rubio, A., Mart{\'{\i}}n-Pintado, J., Thum, C., \& Planesas, P.\ 2013, \aap, 553, AA45 

\bibitem[Beuther et al.(2002)]{beu02}
	Beuther, H., Schilke, P., Sridharan, T.~K., et al.\ 2002, \aap, 383, 892 

\bibitem[Blandford \& Payne(1982)]{bla82}
	Blandford, R.~D., \& Payne, D.~G.\ 1982, \mnras, 199, 883 

\bibitem[Bunn et al.(1995)]{bun95}
	Bunn, J.~C., Hoare, M.~G., \& Drew, J.~E.\ 1995, \mnras, 272, 346 

\bibitem[Butler \& Tan(2012)]{but12}
	Butler, M.~J., \& Tan, J.~C.\ 2012, \apj, 754, 5 

\bibitem[Butler et al.(2014)]{but14}
	Butler, M.~J., Tan, J.~C., \& Kainulainen, J.\ 2014, \apjl, 782, LL30

\bibitem[Carrasco-Gonz{\'a}lez et al.(2010)]{car10}
	Carrasco-Gonz{\'a}lez, C., Rodr{\'{\i}}guez, L.~F., Anglada, G., et al.\ 2010, Science, 330, 1209 

\bibitem[Carrasco-Gonz{\'a}lez et al.(2015)]{car15}
	Carrasco-Gonz{\'a}lez, C., Torrelles, J.~M., Cant{\'o}, J., et al.\ 2015, Science, 348, 114 

\bibitem[Castelli \& Kurucz(2004)]{cas04}
	Castelli, F., \& Kurucz, R.~L.\ 2004, IAU Symp. No 210, Modelling of Stellar Atmospheres, eds. N. Piskunov et al. 2003, poster A20, arXiv:astro-ph/0405087

\bibitem[Codella et al.(2014)]{cod14}
	Codella, C., Cabrit, S., Gueth, F., et al.\ 2014, \aap, 568, L5 

\bibitem[Curiel et al.(2006)]{cur06}
	Curiel, S., Ho, P.~T.~P., Patel, N.~A., et al.\ 2006, \apj, 638, 878 

\bibitem[Dale et al.(2005)]{dal05}
	Dale, J.~E., Bonnell, I.~A., Clarke, C.~J., \& Bate, M.~R.\ 2005, \mnras, 358, 291 

\bibitem[Davies et al.(2011)]{dav11}
	Davies, B., Hoare, M.~G., Lumsden, S.~L., et al.\ 2011, \mnras, 416, 972 

\bibitem[Draine(2011)]{dra11}
	Draine, B.~T.\ 2011, Physics of the Interstellar and Intergalactic Medium by Bruce T.~Draine.~Princeton University Press, 2011.~ISBN: 978-0-691-12214-4

\bibitem[Dupree \& Goldberg(1970)]{dup70}
	Dupree, A.~K., \& Goldberg, L.\ 1970, \araa, 8, 231 

\bibitem[Ferland et al.(2013)]{fer13}
	Ferland, G.~J., Porter, R.~L., van Hoof, P.~A.~M., et al.\ 2013, RMxAA, 49, 137

\bibitem[Figer(2005)]{fig05}
	Figer, D.~F.\ 2005, \nat, 434, 192

\bibitem[Franco-Hern{\'a}ndez \& Rodr{\'{\i}}guez(2004)]{fra04}
	Franco-Hern{\'a}ndez, R., \& Rodr{\'{\i}}guez, L.~F.\ 2004, \apjl, 604, L105 

\bibitem[Galv{\'a}n-Madrid et al.(2008)]{gal08}
	Galv{\'a}n-Madrid, R., Rodr{\'{\i}}guez, L.~F., Ho, P.~T.~P., \& Keto, E.\ 2008, \apjl, 674, L33 

\bibitem[Garay et al.(2007)]{gar07}
	Garay, G., Mardones, D., Bronfman, L., et al.\ 2007, \aap, 463, 217 
	
\bibitem[Garay et al.(1996)]{gar96}
	Garay, G., Ramirez, S., Rodriguez, L.~F., Curiel, S., \& Torrelles, J.~M.\ 1996, \apj, 459, 193

\bibitem[Gibb et al.(2003)]{gib03}
	Gibb, A.~G., Hoare, M.~G., Little, L.~T., \& Wright, M.~C.~H.\ 2003, \mnras, 339, 101

\bibitem[Goodman et al.(1993)]{goo93}
	Goodman, A.~A., Benson, P.~J., Fuller, G.~A., \& Myers, P.~C.\ 1993, \apj, 406, 528 

\bibitem[Guzm{\'a}n et al.(2012)]{guz12}
	Guzm{\'a}n, A.~E., Garay, G., Brooks, K.~J., \& Voronkov, M.~A.\ 2012, \apj, 753, 51
	
\bibitem[Guzm{\'a}n et al.(2014)]{guz14}
	Guzm{\'a}n, A.~E., Garay, G., Rodr{\'{\i}}guez, L.~F., et al.\ 2014, \apj, 796, 117 

\bibitem[Hoare et al.(2007)]{hoa07}
	Hoare, M.~G., Kurtz, S.~E., Lizano, S., Keto, E., \& Hofner, P.\ 2007, Protostars and Planets V, 181 

\bibitem[Hoare \& Franco(2007)]{hoa07b}
	Hoare, M.~G., \& Franco, J.\ 2007, A Volume Honouring John Dyson, Edited by T.W. Hartquist, J. M. Pittard, and S. A. E. G. Falle. Series: Astrophysics and Space Science Proceedings. Springer Dordrecht, 2007, p.61, arXiv:0711.4912 

\bibitem[Hollenbach et al.(1994)]{hol94}
	Hollenbach, D., Johnstone, D., Lizano, S., \& Shu, F.\ 1994, \apj, 428, 654

\bibitem[Hosokawa \& Omukai(2009)]{hos09}
	Hosokawa, T., \& Omukai, K.\ 2009, \apj, 703, 1810 

\bibitem[Hosokawa et al.(2010)]{hos10}
	Hosokawa, T., Yorke, H.~W., \& Omukai, K.\ 2010, \apj, 721, 478 
	
\bibitem[Hosokawa et al.(2011)]{hos11}
	Hosokawa, T., Omukai, K., Yoshida, N., \& Yorke, H.~W.\ 2011, Science, 334, 1250 

\bibitem[Hummer \& Seaton(1963)]{hum63}
	Hummer, D.~G., \& Seaton, M.~J.\ 1963, \mnras, 125, 437 

\bibitem[Keto(2007)]{ket07}
	Keto, E.\ 2007, \apj, 666, 976

\bibitem[Klaassen et al.(2013)]{kla13}
	Klaassen, P.~D., Juhasz, A., Mathews, G.~S., et al.\ 2013, \aap, 555, A73

\bibitem[K\"onigl \& Pudritz(2000)]{kon00}
	K\"onigl, A., \& Pudritz, R.~E.\ 2000, Protostars and Planets IV, 759 

\bibitem[Kratter et al.(2008)]{kra08}
	Kratter, K.~M., Matzner, C.~D., \& Krumholz, M.~R.\ 2008, \apj, 681, 375 

\bibitem[Krumholz et al.(2009)]{kru09}
	Krumholz, M.~R., Klein, R.~I., McKee, C.~F., Offner, S.~S.~R., \& Cunningham, A.~J.\ 2009, Science, 323, 754 

\bibitem[Krumholz et al.(2005)]{kru05}
	Krumholz, M.~R., McKee, C.~F., \& Klein, R.~I.\ 2005, \apjl, 618, L33
	
\bibitem[Krumholz et al.(2007)]{kru07}
	Krumholz, M.~R., Stone, J.~M., \& Gardiner, T.~A.\ 2007, \apj, 671, 518 

\bibitem[Kuiper et al.(2010)]{kui10}
	Kuiper, R., Klahr, H., Beuther, H., \& Henning, T.\ 2010, \apj, 722, 1556 
	
\bibitem[Kuiper et al.(2011)]{kui11}
	Kuiper, R., Klahr, H., Beuther, H., \& Henning, T.\ 2011, \apj, 732, 20 

\bibitem[Kurtz(2002)]{kur02}
	Kurtz, S.\ 2002, Hot Star Workshop III: The Earliest Phases of Massive Star Birth, 267, 81 

\bibitem[Kurtz(2005)]{kur05}
	Kurtz, S.\ 2005, Massive Star Birth: A Crossroads of Astrophysics, 227, 111 

\bibitem[Lanz \& Hubeny(2007)]{lan07}
	Lanz, T., \& Hubeny, I.\ 2007, \apjs, 169, 83 

\bibitem[Li et al.(2012)]{li12}
	Li, J., Wang, J., Gu, Q., Zhang, Z.-y., \& Zheng, X.\ 2012, \apj, 745, 47 

\bibitem[L{\'o}pez-Sepulcre et al.(2009)]{lop09}
	L{\'o}pez-Sepulcre, A., Codella, C., Cesaroni, R., Marcelino, N., \& Walmsley, C.~M.\ 2009, \aap, 499, 811 

\bibitem[Mart{\'{\i}} et al.(1993)]{mar93}
	Mart{\'{\i}}, J., Rodr{\'{\i}}guez, L.~F., \& Reipurth, B.\ 1993, \apj, 416, 208
	
\bibitem[Mart{\'{\i}} et al.(1995)]{mar95}
	Mart{\'{\i}}, J., Rodr{\'{\i}}guez, \& Reipurth, B.\ 1995, \apj, 449, 184

\bibitem[Mart{\'{\i}} et al.(1998)]{mar98}
	Mart{\'{\i}}, J., Rodr{\'{\i}}guez, L.~F., \& Reipurth, B.\ 1998, \apj, 502, 337 

\bibitem[Martins \& Plez(2006)]{mar06}
	Martins, F., \& Plez, B.\ 2006, \aap, 457, 637 

\bibitem[Masqu{\'e} et al.(2015)]{mas15}
	Masqu{\'e}, J.~M., Rodr{\'{\i}}guez, L.~F., Araudo, A., et al.\ 2015, \apj, in press (arXiv:1510.01769)

\bibitem[Matzner \& McKee(2000)]{mat00}
	Matzner, C.~D., \& McKee, C.~F.\ 2000, \apj, 545, 364 

\bibitem[McKee \& Tan(2003)]{mck03}
	McKee, C.~F., \& Tan, J.~C.\ 2003, \apj, 585, 850 (MT03)

\bibitem[McKee \& Tan(2008)]{mck08}
	McKee, C.~F., \& Tan, J.~C.\ 2008, \apj, 681, 771 

\bibitem[McLaughlin \& Pudritz(1997)]{mcl97}
	McLaughlin, D.~E., \& Pudritz, R.~E.\ 1997, \apj, 476, 750 

\bibitem[Meynet et al.(2010)]{mey10}
	Meynet, G., Georgy, C.,  Revaz, Y., et al.\ 2010, Revista Mexicana de Astronomia y Astrofisica Conference Series, 38, 113

\bibitem[Nakano et al.(1995)]{nak95}
	Nakano, T., Hasegawa, T., \& Norman, C.\ 1995, \apj, 450, 183 

\bibitem[Nisini et al.(1994)]{nis94}
	Nisini, B., Smith, H.~A., Fischer, J., \& Geballe, T.~R.\ 1994, \aap, 290, 463 

\bibitem[Palau et al.(2013)]{pal13}
	Palau, A., Fuente, A., Girart, J.~M., et al.\ 2013, \apj, 762, 120 

\bibitem[Panagia \& Felli(1975)]{pan75}
	Panagia, N., \& Felli, M.\ 1975, \aap, 39, 1 

\bibitem[Peters et al.(2010)]{pet10}
	Peters, T., Klessen, R.~S., Mac Low, M.-M., \& Banerjee, R.\ 2010, \apj, 725, 134
	
\bibitem[Peters et al.(2011)]{pet11}
	Peters, T., Banerjee, R., Klessen, R.~S., \& Mac Low, M.-M.\ 2011, \apj, 729, 72 

\bibitem[Reynolds(1986)]{rey86}
	Reynolds, S.~P.\ 1986, \apj, 304, 713

\bibitem[Richling \& Yorke(1997)]{ric97}
	Richling, S., \& Yorke, H.~W.\ 1997, \aap, 327, 317 

\bibitem[Rodr{\'{\i}}guez et al.(1989a)]{rod89a}
	Rodr{\'{\i}}guez, L.~F., Curiel, S., Moran, J.~M., et al.\ 1989a, \apjl, 346, L85 

\bibitem[Rodr{\'{\i}}guez et al.(2005)]{rod05}
	Rodr{\'{\i}}guez, L.~F., Garay, G., Brooks, K.~J., \& Mardones, D.\ 2005, \apj, 626, 953 

\bibitem[Rodr{\'{\i}}guez et al.(2007)]{rod07} 
	Rodr{\'{\i}}guez, L.~F., G{\'o}mez, Y., \& Tafoya, D.\ 2007, \apj, 663, 1083 

\bibitem[Rodr{\'{\i}}guez et al.(2008)]{rod08}
	Rodr{\'{\i}}guez, L.~F., Moran, J.~M., Franco-Hern{\'a}ndez, R., et al.\ 2008, \aj, 135, 2370

\bibitem[Rodr{\'{\i}}guez et al.(1989b)]{rod89}
	Rodr{\'{\i}}guez, L.~F., Myers, P.~C., Cruz-Gonzalez, I., \& Terebey, S.\ 1989b, \apj, 347, 461 

\bibitem[Salem \& Brocklehurst(1979)]{sal79}
	Salem, M., \& Brocklehurst, M.\ 1979, \apjs, 39, 633

\bibitem[Shakura \& Sunyaev(1973)]{sha73}
	Shakura, N.~I., \& Sunyaev, R.~A.\ 1973, \aap, 24, 337 

\bibitem[Shu(1977)]{shu77}
	Shu, F.~H.\ 1977, \apj, 214, 488 
	
\bibitem[Stone et al.(1992)]{sto92}
	Stone, J.~M., Mihalas, D., \& Norman, M.~L.\ 1992, \apjs, 80, 819

\bibitem[Strelnitski et al.(1996)]{Str96}
	Strelnitski, V.~S., Ponomarev, V.~O., \& Smith, H.~A.\ 1996, \apj, 470, 1118 

\bibitem[Tafoya et al.(2004)]{taf04}
	Tafoya, D., G{\'o}mez, Y., \& Rodr{\'{\i}}guez, L.~F.\ 2004, \apj, 610, 827 

\bibitem[Tan \& McKee(2003)]{tan03}
	Tan, J.~C., \& McKee, C.~F.\ 2003, IAU Symposium 221, Star Formation at High Angular Resolution, Eds. M. Burton, R. Jayawardhana, \& T. Bourke, arXiv:astro-ph/0309139

\bibitem[Tan \& McKee(2004)]{tan04}
	Tan, J.~C., \& McKee, C.~F.\ 2004, \apj, 603, 383 

\bibitem[Tan \& McKee(2004)]{tan04b}
	Tan, J.~C., \& McKee, C.~F.\ 2004, The Formation and Evolution of Massive Young Star Clusters, 322, 263 

\bibitem[Tan et al.(2014)]{tan14}
	Tan, J.~C., Beltr{\'a}n, M.~T., Caselli, P., et al.\ 2014, Protostars and Planets VI, 149 

\bibitem[Tanaka \& Nakamoto(2011)]{tan11}
	Tanaka, K.~E.~I., \& Nakamoto, T.\ 2011, \apjl, 739, L50 

\bibitem[Tanaka et al.(2013)]{tan13}
	Tanaka, K.~E.~I., Nakamoto, T., \& Omukai, K.\ 2013, \apj, 773, 155

\bibitem[Thompson(1984)]{tho84}
	Thompson, R.~I.\ 1984, \apj, 283, 165 

\bibitem[Thum et al.(1994)]{thu94}
	Thum, C., Matthews, H.~E., Martin-Pintado, J., et al.\ 1994, \aap, 283, 582

\bibitem[Ulrich(1976)]{ulr76}
	Ulrich, R.~K.\ 1976, \apj, 210, 377

\bibitem[Walmsley(1990)]{wal90}
	Walmsley, C.~M.\ 1990, \aaps, 82, 201 

\bibitem[Weingartner \& Draine(2001)]{wei01}
	Weingartner, J.~C., \& Draine, B.~T.\ 2001, \apj, 548, 296

\bibitem[Wilner et al.(1999)]{wil99}
	Wilner, D.~J., Reid, M.~J., \& Menten, K.~M.\ 1999, \apj, 513, 775 

\bibitem[Wood \& Churchwell(1989)]{woo89}
	Wood, D.~O.~S., \& Churchwell, E.\ 1989, \apjs, 69, 831 

\bibitem[Yorke \& Bodenheimer(1999)]{yor99}
	Yorke, H.~W., \& Bodenheimer, P.\ 1999, \apj, 525, 330 

\bibitem[Yorke \& Welz(1996)]{yor96}
	Yorke, H.~W., \& Welz, A.\ 1996, \aap, 315, 555 

\bibitem[Zhang \& Tan(2011)]{zha11}
	Zhang, Y., \& Tan, J.~C.\ 2011, \apj, 733, 55 

\bibitem[Zhang et al.(2013)]{zha13}
	Zhang, Y., Tan, J.~C., \& McKee, C.~F.\ 2013, \apj, 766, 86 

\bibitem[Zhang et al.(2014)]{zha14}
	Zhang, Y., Tan, J.~C., \& Hosokawa, T.\ 2014, \apj, 788, 166 (ZTH14)


\end{thebibliography}
\end{document}